\documentclass[11pt, letterpaper, twocolumn, logo]{miulab}

\PassOptionsToPackage{comma,numbers,sort,compress}{natbib}

\usepackage[comma,numbers,sort,compress]{natbib}

\usepackage{hyperref}
\usepackage{gdm-colors}
\usepackage{enumitem}
\usepackage{cleveref}
\usepackage{svg}
\usepackage{makecell}
\usepackage[utf8]{inputenc}
\usepackage[T1]{fontenc}
\hypersetup{
    colorlinks = true,
    citecolor = {magenta},
}

\usepackage{lipsum}

\newcommand\blfootnote[1]{%
  \begingroup
  \renewcommand\thefootnote{}\footnote{#1}%
  \addtocounter{footnote}{-1}%
  \endgroup
}

\pdftrailerid{redacted}

\makeatletter
\renewcommand\bibentry[1]{\nocite{#1}{\frenchspacing\@nameuse{BR@r@#1\@extra@b@citeb}}}
\makeatother

\usepackage{kantlipsum, lipsum}
\usepackage{gdm-colors}
\usepackage{subfigure}
\usepackage{wrapfig}
\usepackage{multirow}
\usepackage{graphicx}
\usepackage{bm}
\usepackage{mathrsfs}
\usepackage{float}
\usepackage[section]{placeins}

\usepackage[most,skins,theorems]{tcolorbox}
\tcbuselibrary{breakable}

\tcbset{
  aibox/.style={
    width=\linewidth,
    top=8pt,
    bottom=4pt,
    colback=blue!6!white,
    colframe=black,
    colbacktitle=black,
    enhanced,
    center,
    attach boxed title to top left={yshift=-0.1in,xshift=0.15in},
    boxed title style={boxrule=0pt,colframe=white,},
  }
}
\newtcolorbox{AIbox}[2][]{aibox,title=#2,#1}
\definecolor{lightblue}{rgb}{0.22,0.45,0.70}

\usepackage{algorithm}
\usepackage{algpseudocode}

\DeclareFloatingEnvironment{boxes} 
\newcommand{\boxref}[1]{\hyperref[{#1}]{TextBox~\ref*{#1}}}

\graphicspath{{figures/}}

\title{\texorpdfstring{\includegraphics[width=0.8cm]{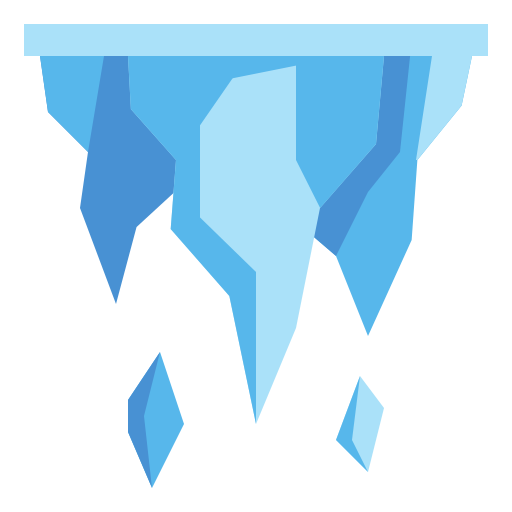}}{} ICICLE: Expanding Retrieval with In-Context Documents}

\author[1,$*$]{Yu-Chen Den}
  \author[1,$*$]{Yung-Yu Shih}
  \author[1]{Zhi Rui Tam}
  \author[1]{Kuan-Yu Chen}
  \author[1]{Pu-Jen Cheng}
  \author[1]{Yun-Nung Chen}
  \author[2]{Eugene Yang}

  \affil[1]{National Taiwan University, Taipei, Taiwan}
  \affil[2]{Johns Hopkins University, Baltimore, US}
\begin{abstract}
{\centering
  \vspace{-0.5em}
  \includegraphics[height=1.2em]{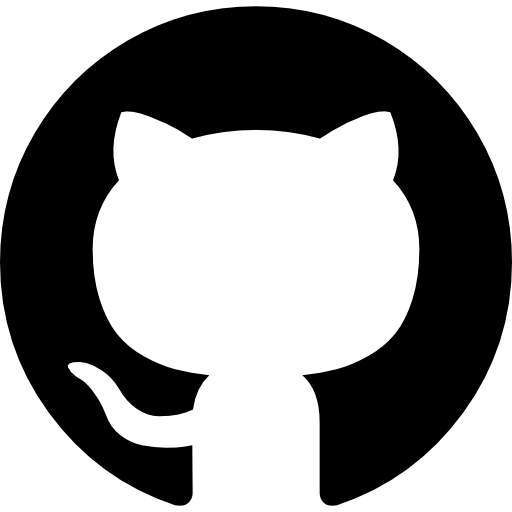}~
  \href{}{\texttt{Code}}~~
  \includegraphics[height=1.2em]{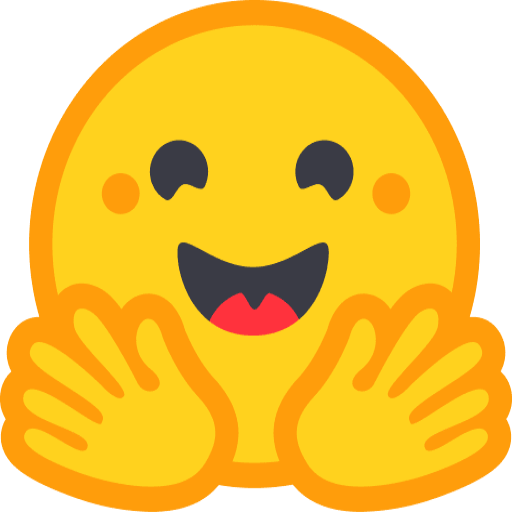}~
  \href{}{\texttt{Models}}~~
  \includegraphics[height=1.2em]{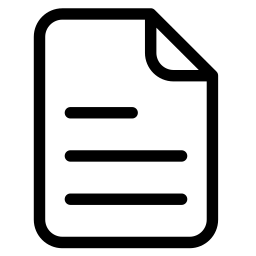}~
  \href{}{\texttt{Project Page}}
\par
}
\vspace{7mm}
{
\centering
\par
}

Generative retrieval (GR) maps queries directly to document identifiers (docids) using parametric knowledge, However, this design makes corpus expansion costly: adding new documents requires updating model parameters to encode new document-docid associations incurs repeated training and catastrophic forgetting of previously indexed documents. In this work, we revisit incremental GR as an in-context retrieval problem, where newly added documents are supplied as inference-time document-docid evidence. We propose ICICLE, an in-context indexing framework that performs source-aware docid generation over both parametric memory and context-provided document-docid pairs. ICICLE combines a \texttt{[COPY]}-based routing mechanism, preference-based calibration, and large context adaptation to distinguish context-grounded retrieval from parametric retrieval. Experiments on MS MARCO and NQ320K show that ICICLE improves retrieval of newly introduced documents while preserving seen-document retention without corpus-specific retraining. Our analysis further shows that high-shot degradation is mainly caused by routing failure, highlighting source-selection calibration as a key bottleneck for scaling in-context generative retrieval.
\end{abstract}
\begin{document}
\maketitle
\blfootnote{$^*$Equal contribution}

\section{Introduction}
Generative retrieval (GR) has emerged as a rapidly growing paradigm for information retrieval, demonstrating strong performance across diverse tasks—including document retrieval, entity search, and recommendation—while outperforming dense retrieval baselines across settings~\citep{tay2022transformer, wang2022neural, rajput2023recommender}. Unlike dense retrievers, which encode documents into vectors stored in an external index, GR trains a language model to directly map queries to document identifiers (docids), encoding document–query associations into model parameters. This parametric design eliminates the need for an external document index, reducing both the memory overhead of storing large-scale document embeddings and the query-time cost of approximate nearest neighbor search~\citep{sun2023learning}. However, real-world retrieval systems typically contend with evolving corpora, where new documents are continuously added after deployment~\citep{lawrie2024plaid,kishore2023incdsi}. While dense retrievers accommodate this naturally by inserting new vectors into the index~\citep{karpukhin2020dense,chen2024m3}, GR faces a fundamental tension: retrieving a document requires its content–docid association to be encoded in model parameters~\citep{zhang2025replication}, making the addition of new documents a model-update problem rather than a simple indexing operation~\citep{mehta2023dsi++,guo2025corpusbrain++}.

Existing approaches address this by incorporating new documents through continued training, parameter-efficient expansion, constrained parameter updates, or model editing, all of which modify model parameters to encode newly added document-docid associations~\citep{mehta2023dsi++,kishore2023incdsi,huynh2025mixlora,zhang2026model,son2026cream}.
While these methods can extend the retrievable corpus, they tie corpus growth to repeated optimization. Each update may require additional training, validation, and redeployment, and repeated modifications can interfere with previously learned docid mappings, leading to catastrophic forgetting of old documents and higher operational costs~\citep{mehta2023dsi++,mekonnen2026parametric}. 
This makes it difficult to deploy GR in settings where corpora evolve frequently. We therefore ask: Can a GR model retrieve newly added, context-provided documents while preserving retrieval over previously learned documents, without document-specific parameter updates whenever the corpus grows?

Unlike existing approaches that update model parameters to incorporate new document–docid associations, we expose them in the input context at inference time~\citep{ram2023context}, casting incremental GR as an \textit{in-context retrieval} problem: given a query and a set of context-provided document–docid pairs, the model must generate the docid of the relevant document. In contrast to standard in-context learning (ICL), where demonstrations specify how to solve a task~\citep{brown2020language,dong2024survey}, in-context retrieval uses document–docid pairs as a temporary index over newly added documents—requiring the model to combine parametric memory for seen documents with contextual evidence for unseen ones.

However, simply prompting a GR model with document–docid pairs is insufficient. The model must determine whether a query should be answered from parametric memory or from the context-provided index. When parametric memory dominates, the model may return a memorized docid even if the relevant new document is in context~\citep{lee2023glen}; when contextual cues dominate, it may latch onto an irrelevant in-context docid due to shallow prompt artifacts~\citep{min2022rethinking,shi2024trusting,xu2024knowledge}. Both failure modes result in incorrect docid generation.

To address this, we introduce \emph{\textbf{I}n-\textbf{C}ontext \textbf{I}ndexing for do\textbf{C}-\textbf{L}evel \textbf{E}xpansion (ICICLE)}, a framework that trains GR models to perform source-aware docid generation via explicit routing supervision and preference-based calibration. ICICLE encourages the model to ground generation in context when the relevant document is newly provided, while preserving parametric retrieval for seen documents—extending GR from a closed-corpus retriever to one that operates over dynamically supplied documents without parameter updates.

Our main contributions are as follows:
\begin{itemize}
    \item We cast incremental GR as \textit{in-context retrieval}, a new problem formulation in which newly added documents are supplied as inference-time document–docid evidence, eliminating the need for document-specific parameter updates.
    \item We introduce ICICLE, a three-stage training framework for source-aware docid generation. ICICLE combines a \texttt{[COPY]}-based routing mechanism with DPO-based calibration to distinguish context-grounded from parametric retrieval, and a long-context adaptation stage to bridge the train-test mismatch under large candidate pools.
    \item We conduct experiments on MS MARCO~\citep{NguyenRSGTMD16} and NQ320K~\citep{47761}, showing that ICICLE is a strong alternative to incremental GR baselines, improving retrieval of newly introduced documents while preserving retention on seen documents without corpus-specific retraining. Further analysis shows that high-shot degradation mainly stems from routing failure, highlighting source-selection calibration as a key bottleneck for scaling in-context generative retrieval.
\end{itemize}
\section{Preliminaries}
\subsection{Generative Retrieval}

Given a document corpus $\mathcal{D}=\{d_1,d_2,\ldots,d_N\}$, generative retrieval (GR) retrieves documents by directly generating their identifiers. For a query $q$, a GR model autoregressively generates a document identifier (docid) $\bm{y}=(y_1,y_2,\ldots,y_T)$:
\begin{equation}
    p_\theta(\bm{y}\mid q)=\prod_{t=1}^{T} p_\theta(y_t \mid y_{<t}, q),
\end{equation}
where $\theta$ denotes the model parameters.
Following prior work~\citep{tay2022transformer}, GR is typically trained with both retrieval and indexing objectives: the model learns to generate the docid from either a relevant query or the document content itself. Given a query--document pair $(q,d)$ with docid $\bm{y}$, the training loss is:
\begin{equation}
    \scalebox{0.85}{$\mathscr{L}_{\mathrm{GR}} = -\displaystyle\sum_{t=1}^{T} \log p_\theta(y_t \mid y_{<t}, q) -\displaystyle\sum_{t=1}^{T} \log p_\theta(y_t \mid y_{<t}, d)$}.
\end{equation}
In this work, we adopt the \textit{title-as-docid} scheme~\citep{de2020autoregressive,lee2023nonparametric}, where each document is identified by its title. This avoids introducing an auxiliary docid vocabulary and naturally assigns identifiers to newly arriving documents without modifying the model's token space. At inference time, the model generates a docid using constrained beam search over valid docids, and the associated document is returned as the retrieval result.

\subsection{Incremental Learning for Generative Retrieval}
\label{subsec:inc_learning}

Standard GR assumes a static corpus fixed at training time. In real-world applications, however, document collections are often dynamic: new documents may be created or ingested after the retriever has already been trained and deployed.

We formalize this setting as follows. Let $\mathcal{D}_{\mathrm{train}}=\{d_1,\ldots,d_N\}$ denote the corpus used to train a GR model $f_\theta$. After deployment, a set of unseen documents $\mathcal{D}_{\mathrm{new}}=\{d_{N+1},\ldots,d_{N+M}\}$ may be added, where $\mathcal{D}_{\mathrm{new}}\cap\mathcal{D}_{\mathrm{train}}=\emptyset$. The retriever is then expected to answer queries over the expanded corpus $\mathcal{D}_{\mathrm{train}}\cup\mathcal{D}_{\mathrm{new}}$.

This setting is challenging because documents in $\mathcal{D}_{\mathrm{new}}$ are absent during training, the model has not learned the document---docid associations. Although the underlying language model may understand an unseen document's content, GR requires generating it's exact docid. Therefore, for a query \(q\) whose relevant document \(d^+\in\mathcal{D}_{\mathrm{new}}\), the original model \(f_\theta\) may fail unless the new document--docid association is retrained into the model or otherwise provided at inference time.
\section{ICICLE: In-Context Indexing for doC-Level Expansion}
\label{sec:method}

\begin{figure*}[t]
    \centering
    \includegraphics[width=0.8\textwidth]{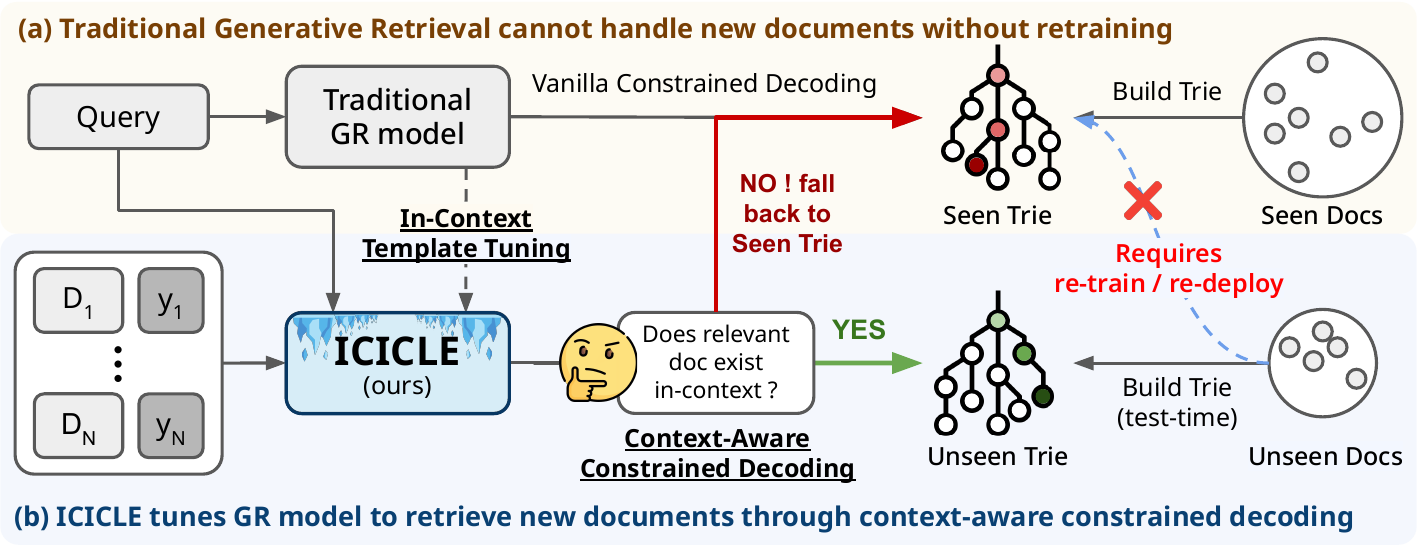}
    \caption{Overview of ICICLE. (a)~Traditional GR models are constrained to a fixed trie built from seen documents and cannot handle unseen documents without retraining. (b)~ICICLE tunes the GR model with in-context template to perform context-aware constraint decoding: given unseen candidate document-docid pairs at test time, the model decides whether the relevant document exists in context --- if so, it decodes against a dynamically constructed unseen trie; otherwise, it falls back to its own parametric memory without any retraining.}
    \label{fig:icicle}
\end{figure*}

Building on the incremental GR setting introduced in \Cref{subsec:inc_learning}, ICICLE aims to support retrieval over newly added documents without retraining the generative retriever whenever the corpus expands. The key idea is to treat the documents provided in the prompt as a temporary in-context index: instead of encoding every new document into the model parameters, ICICLE learns to retrieve from document-docid pairs supplied at inference time. As shown in~\Cref{fig:icicle}, ICICLE first learns an in-context retrieval template, then calibrates whether the model should retrieve from the contextual index or fall back to parametric memory. At inference time, newly added documents can be incorporated by updating only the prompt-side candidates and the dynamically constructed decoding trie.

\subsection{Learning In-Context Template}
\label{subsec:icl_template}

Given a base generative retriever \(f_{\theta}\), we finetune it to condition generation on a small set of in-context document-docid pairs. 
For each training query \(q \in \mathcal{Q}_{\text{train}}\) with gold document \(d^{*}\) and gold docid \(y^{*}\), we construct an in-context candidate set
\begin{equation}
    \mathcal{C}=\{(d_{1}, y_{1}), \ldots, (d_{n}, y_{n})\}.
\end{equation}
The model is trained to generate \(y^{*}\) conditioned on \((q, \mathcal{C})\). To teach the model when to use the context and when to ignore it, we construct two types of training instances: 

\paragraph{Context-Dependent.}
The gold pair \((d^*, y^*)\) is inserted into \(\mathcal{C}\) at a uniformly sampled position, while the remaining \(n-1\) positions are filled with hard negative documents \(\mathcal{H}(d^{*})\) retrieved from a dense embedding space as semantic neighbors of \(d^{*}\), making the discrimination non-trivial. The model is expected to uses the provided candidate set as a temporary index and copies the corresponding docid from the context.

\paragraph{Query-Irrelevant.}
The entire context contains only hard negatives drawn from \(\mathcal{H}(d^{*})\) and excludes \((d^*, y^*)\). Here, \(y^{\star}\) belongs to the base corpus known to the original GR model, so the model should ignore the irrelevant context and fall back to parametric retrieval. These instances prevent the model from over-copying irrelevant contextual docids when the answer is absent. \\

This design reflects real-world retrieval settings in which the answer to an incoming query may or may not be present in the provided context. Therefore, the model must learn not only to retrieve from contextual evidence when it is relevant, but also to ignore irrelevant context and fall back to parametric generative retrieval when necessary. We show examples of the full in-context template in Appendix~\ref{appendix:icl_template}.

\paragraph{Document Compression.}
To fit more candidates into the prompt, we compress each document offline before constructing in-context instances. Each document is summarized by Qwen2.5-14B-Instruct~\citep{yang2025qwen3} into a retrieval-oriented representation of at most \(256\) tokens, preserving key entities, dates, aliases, relations, and facts. 
The compression model only observes document content and does not access queries, relevance labels, or test annotations. At inference time, each candidate is represented by its compressed text followed by its title docid. The full compression analysis is provide in \Cref{appendix:compression_comparison}. 

\subsection{Routing Between Contextual and Parametric Retrieval}
A key challenge is to calibrate whether the model should ground its prediction in the provided context or rely on its parametric memory. This distinction is non-trivial because contextual identifiers and memorized identifiers share the same output space, while standard sequence likelihood training only supervises the final identifier and does not explicitly indicate which retrieval mode should be used. We address this issue by introducing a \texttt{[COPY]} router token and further calibrating the routing behavior with preference optimization.

\paragraph{\texttt{[COPY]} Token as Router.} To distinguish context-based retrieval from parametric recall, we introduce a special token \texttt{[COPY]}. As described in~\Cref{subsec:icl_template}, the gold document may either be present in the provided candidate set or be absent from it. Without an explicit signal, the model must use the same output format for both cases, making it unclear whether the docid should be retrieved from the context or generated from its own memory.

During training, if the gold pair appears in the context, the model is trained to output \texttt{[COPY]} token followed by the docid. Otherwise, it directly generates the plain docid. Formally, the supervision target is
\begin{equation}
    \tilde{y}_i =
    \begin{cases}
    \texttt{[COPY]}\; y^{*}, & \text{if } y^{*} \in \mathcal{Y}(\mathcal{C}),\\
    y^{*}, & \text{otherwise}.
    \end{cases}
\end{equation}
We optimize the standard next-token cross-entropy objective:
\begin{equation}
    \mathscr{L}_{\text{SFT}} = -\sum_{i}\log p_{\theta}\left(\tilde{y}_i \,\bigm|\, \mathcal{C}_i,\, q_i\right).
\end{equation}
At inference time, \texttt{[COPY]} determines the constrained decoding space.  If the model generates \texttt{[COPY]} as the first token, subsequent decoding is constrained by a trie dynamically built from the in-context docids \(\mathcal{Y}(\mathcal{C}_q)\); otherwise, decoding uses the base trie over \(\mathcal{Y}_{\mathrm{base}}\). The \texttt{[COPY]} token is used only as an internal routing signal and is removed before evaluation. Details are provided in Appendix~\ref{appendix:constrained_decoding}

\paragraph{Calibration with DPO.}
Although SFT on in-context pairs teaches the model the desired output format, it does not explicitly calibrate the routing decision induced by the \texttt{[COPY]} token.  We observe that an SFT model may learn the template while still failing to activate this router reliably at test time, either by ignoring the context and falling back to parametric retrieval, or by ranking plausible but incorrect identifiers above the gold one during beam search.

We therefore apply Direct Preference Optimization (DPO)~\citep{rafailov2023direct} after SFT. For each prompt \((\mathcal{C}, q)\), we construct a chosen response \(y^{+}\) and a rejected response \(y^{-}\). We consider two types of preference pairs. First for ranking failures, if the gold docid appears in the top-$B$ beams candidates but is not ranked first, we use the gold docid as \(y^{+}\) and the top-ranked incorrect beam as \(y^{-}\). Second, for routing failures in context-dependent samples, we use \texttt{[COPY] $y^*$} as the \(y^{+}\) and an incorrect zero-shot context-blind prediction as \(y^{-}\). Although the rejected docid is mined from zero-shot or context-blind predictions, both the chosen and rejected responses are paired with the same in-context prompt during DPO optimization. The detail of DPO is provided in Appendix~\ref{appendix:dpo}.

\subsection{Large Title Context Adaptation}
\label{subsec:lora}

SFT and DPO stages use short document-docid contexts due to input length constraints. However, at inference time, we may involve up to $K=100$ in-context pairs. This creates a train-test mismatch: the model is optimized with short, context-rich demonstrations but evaluated with larger candidate lists. Moreover, including full document texts for all \(K\) candidates is often infeasible under a fixed context budget. To bridge this mismatch, we add a lightweight large context adaptation stage using title-only context:
\begin{equation}
    \hat{\mathcal{C}}=\{y_i\}_{i=1}^{K},
\end{equation}
where each \(y_i\) is a natural-language title docid. Since titles preserve lexical and semantic cues, this provides a compact approximation of a long candidate index. 

Starting from the DPO-trained model, we freeze the backbone and update only LoRA parameters:
\begin{equation}
    \max_{\Delta\theta_{\mathrm{LoRA}}}
    \log p_{\theta+\Delta\theta_{\mathrm{LoRA}}}
    (y^\ast \mid q, \hat{\mathcal{C}}).
\end{equation}
The backbone $\theta$ is frozen and only the LoRA parameters are updated. This stage adapts the model to larger title-level prompts following the idea of efficient context extension~\citep{yang2023longqlora}. This adaptation is performed only on training corpus instances and does not use test queries, relevance labels, or document-query pairs from the evaluation set. Details are provided in Appendix~\ref{app:context_adaptation}.




\section{Experiments}

\subsection{Experiment Settings}
\label{subsec:exp_setting}

\paragraph{Datasets.} We conduct experiments on Natural Questions 320K (NQ320K)~\citep{47761} and MS MARCO~\citep{NguyenRSGTMD16}\footnote{We use the smaller v1.1 dataset due to restricted computational resource}. 
To simulate the arrival of new documents, we split each corpus into an initial corpus \(\mathcal{D}_{\text{train}}\) and an unseen corpus \(\mathcal{D}_{\text{new}}\), containing 90\% and 10\% of the documents, respectively. The base GR model is trained only on \(\mathcal{D}_{\text{train}}\), while \(\mathcal{D}_{\text{new}}\) is introduced only at adaptation and evaluation time. We evaluate adaptation using queries targeting \(\mathcal{D}_{\text{new}}\), and retention using queries targeting \(\mathcal{D}_{\text{train}}\). Further details are provided in Appendix~\ref{sec:data}.

\paragraph{Baselines.}
We compare ICICLE with two groups of baselines, distinguished by whether they are explicitly adapted to the unseen corpus \(\mathcal{D}_{\text{new}}\). The first group is not trained or adapted on \(\mathcal{D}_{\text{new}}\). It includes the original DSI model~\citep{tay2022transformer}, which is trained only on \(\mathcal{D}_{\text{train}}\) but decodes over a trie containing document ids from both \(\mathcal{D}_{\text{train}}\) and \(\mathcal{D}_{\text{new}}\). Thus, its retrieval performance on \(\mathcal{D}_{\text{new}}\) reflects the model's parametric generalization ability rather than explicit access to new documents.

The second group has access to \(\mathcal{D}_{\text{new}}\) through direct indexing, retraining, fine-tuning, or model editing. This includes sparse retriever BM25~\citep{robertson2009probabilistic} and dense retriever DPR~\citep{karpukhin2020dense}, which directly index new documents; \textsc{From Scratch}, which trains a GR model on \(\mathcal{D}_{\text{train}} \cup \mathcal{D}_{\text{new}}\); New-Doc FT and DSI++~\citep{mehta2023dsi++}, which incrementally fine-tune the GR model with new-document data; and DOME~\citep{zhang2026model}, which adapts GR models through model editing.

\paragraph{Evaluation Protocol.}
ICICLE retrieves documents from two sources: retrieve from its parametric memory over \(\mathcal{D}_{\text{train}}\), while the in-context document-docid pairs provide query-time access to newly added documents from \(\mathcal{D}_{\text{new}}\). For each unseen query with gold document \(d^+ \in \mathcal{D}_{\text{new}}\), we construct a query-specific unseen candidate set \(D_q\) by combining \(d^+\) with \(N-1\) randomly sampled non-gold documents:
\begin{equation}
    D_q = \{d^+\} \cup \{d_i^-\}_{i=1}^{N-1}, 
    \  d_i^- \in \mathcal{D}_{\text{new}} \setminus \{d^+\}.
\end{equation}
Since the number of in-context pairs is bounded by the maximum context length, we set \(N=100\) in the main experiments. ICICLE receives \(D_q\) as in-context document-docid pairs and retrieves over \(\mathcal{D}_{\text{train}} \cup D_q\). For non-context baselines, we build the corresponding retrieval index or decoding trie over the same search space \(\mathcal{D}_{\text{train}} \cup D_q\). Therefore, all methods are evaluated under the same query-specific search space, and differ only in how they incorporate newly added documents.

\paragraph{Evaluation Metrics.}
Following prior work on GR~\citep{tay2022transformer,mehta2023dsi++}, we report Hits@1 and Hits@10. We report adaptation performance on queries targeting \(\mathcal{D}_{\text{new}}\), and retention performance on seen-document queries targeting \(\mathcal{D}_{\text{train}}\). Efficiency is measured as the theoretical training cost required to incorporate unseen  documents into the model weights, where $N$ and $M$ denote the size of $\mathcal{D}_{\text{train}}$ and $\mathcal{D}_{\text{new}}$, respectively.

\paragraph{Implementation Details.}
Unless otherwise specified, ICICLE uses \(N=100\) in-context document-docid pairs. All GR-based methods use the same document-id space and decoding constraints for fair comparison. Training hyperparameters, prompt templates, and baseline-specific configurations are provided in Appendix~\ref{appendix:training_conf}.

\subsection{Main Result}
\label{subsec:main_result}

\begin{table*}[t]
    \centering
    \caption{
        Main results on MS MARCO and NQ320K for dynamic corpus retrieval. $\mathcal{D}_{\text{train}}$ is the initially indexed corpus and $\mathcal{D}_{\text{new}}$ is the unseen corpus added at inference time. We evaluate retrieval performance on both corpora using Hits@1 and Hits@10, and compare the theoretical training cost required to support the expanded corpus $\mathcal{D}_{\text{new}}$. \textbf{Bold} and \underline{underline} shows best and second best result, respectively. BM25 \& DPR are non-GR references, while the main comparison is among GR methods.
    }
    \label{tab:main_result}
    \resizebox{0.9\textwidth}{!}{%
    \begin{tabular}{lccccccccc}
    \toprule
    & \multicolumn{4}{c}{\textbf{MS MARCO}}
    & \multicolumn{4}{c}{\textbf{NQ320K}}
    & \multirow{3}{*}[-1.5ex]{\textbf{\shortstack[c]{Unseen Corpus\\Update Cost}}} \\
    \cmidrule(lr){2-5} \cmidrule(lr){6-9}
    \textbf{Method}
    & \multicolumn{2}{c}{\textbf{Eval on $\mathcal{D}_{\text{train}}$}}
    & \multicolumn{2}{c}{\textbf{Eval on $\mathcal{D}_{\text{new}}$}}
    & \multicolumn{2}{c}{\textbf{Eval on $\mathcal{D}_{\text{train}}$}}
    & \multicolumn{2}{c}{\textbf{Eval on $\mathcal{D}_{\text{new}}$}} & \\
    \cmidrule(lr){2-3} \cmidrule(lr){4-5} \cmidrule(lr){6-7} \cmidrule(lr){8-9}
    & \textbf{H@1} & \textbf{H@10}
    & \textbf{H@1} & \textbf{H@10}
    & \textbf{H@1} & \textbf{H@10}
    & \textbf{H@1} & \textbf{H@10} & \\
    \midrule
    \multicolumn{9}{l}{\textbf{\textit{Indexed on \(\mathcal{D}_{\text{train}}\)}}} \\
    DSI & 0.634 & 0.850 & 0.003 & 0.006 & 0.424 & 0.608 & 0.156 & 0.317 & -- \\
    \midrule
    \multicolumn{9}{l}{\textbf{\textit{Indexed on \(\mathcal{D}_{\text{train}} \cup \mathcal{D}_{\text{new}}\)}}} \\
    BM25$^*$ & 0.614 & 0.937 & 0.601 & 0.928 & 0.386 & 0.617 & 0.297 & 0.471 & -- \\
    DPR$^*$ & 0.444 & 0.752 & 0.477 & 0.779 & 0.471 & 0.719 & 0.479 & 0.723 & -- \\
    From Scratch & \textbf{0.590} & \textbf{0.813} & 0.603 & 0.788 & \underline{0.450} & \underline{0.611} & 0.395 & 0.559 & \(\mathcal{O}(N+M)\) \\
    New-Doc FT & 0.437 & 0.690 & \textbf{0.676} & \textbf{0.836} & 0.418 & 0.570 & 0.432 & 0.556 & \(\mathcal{O}(M)\) \\
    DSI++ & 0.485 & 0.696 & 0.506 & 0.666 & 0.422 & 0.607 & 0.604 & 0.723 & \(\mathcal{O}(M)\) \\
    DOME & 0.450 & 0.606 & 0.518 & 0.692 & 0.424 & 0.608 & \underline{0.609} & \underline{0.725} & \(\mathcal{O}(M)\) \\
    \textbf{ICICLE (Ours)}
    & \underline{0.519} & \underline{0.772} & \underline{0.607} & \underline{0.800}
    & \textbf{0.483} & \textbf{0.664} & \textbf{0.649} & \textbf{0.772} & -- \\

    \bottomrule
    \end{tabular}
    }
\end{table*}
\Cref{tab:main_result} reports retrieval performance on both the originally indexed corpus \(\mathcal{D}_{\text{train}}\) and the newly added corpus \(\mathcal{D}_{\text{new}}\). 
We include BM25 and DPR as non-parametric retrieval references indexed over \(\mathcal{D}_{\text{train}} \cup \mathcal{D}_{\text{new}}\), while focusing our main comparison on GR-based incremental methods. Overall, ICICLE achieves strong incremental retrieval performance without any document-specific parameter update. On \(\mathcal{D}_{\text{new}}\), ICICLE obtains 0.607 Hits@1 and 0.800 Hits@10 on MS MARCO, outperforming DSI++ and DOME and approaching oracle-style methods such as From Scratch and New-Doc FT that directly train on new-document supervision. On NQ320K, ICICLE further achieves the best performance on newly added documents, with 0.649 Hits@1 and 0.772 Hits@10. Meanwhile, ICICLE preserves competitive performance on \(\mathcal{D}_{\text{train}}\), including the strongest Hits@1 among GR-based methods on NQ320K. These results show that in-context document-docid pairs provide an effective lightweight interface for corpus expansion, shifting incremental adaptation from offline model updates to query-time context conditioning.

Beyond retrieval, ICICLE substantially reduces the cost of supporting an expanded corpus. For standard generative retrieval methods, unseen documents cannot be retrieved unless their identifiers are incorporated into the model through additional training. As shown in Table~\ref{tab:main_result}, retraining from scratch requires optimization over the full expanded corpus \(\mathcal{D}_{\mathrm{train}} \cup \mathcal{D}_{\mathrm{new}}\), resulting in an update cost of \(\mathcal{O}(N+M)\), while continual fine-tuning methods still require updating the model on the newly added \(O(M)\) documents. By contrast, ICICLE performs document expansion non-parametrically: after a one-time training stage that teaches the model to use in-context document-docid pairs, new documents can be added at inference time by updating only the prompt-side candidate set and the dynamic decoding trie. Therefore, ICICLE does not need to retrain or fine-tune the model whenever unseen documents are introduced, making it more efficient for dynamic retrieval settings. Moreover, we provide inference latency of different number of shots in Appendix~\ref{appendix:latency}.

\subsection{Ablation Studies}
\label{subsec:ablation}

\Cref{tab:ablation} evaluates each component of ICICLE on \(\mathcal{D}_{\text{train}}\) and \(\mathcal{D}_{\text{new}}\). Direct inference performs poorly, showing that instruction following alone is insufficient for mapping newly provided documents to docids. The ICL template improves retrieval, but the \texttt{[COPY]} token brings the largest gain on \(\mathcal{D}_{\text{new}}\), increasing Hits@1 from 0.102 to 0.526 and confirming the need for an explicit context-copying signal. DPO then improves retention on \(\mathcal{D}_{\text{train}}\), raising Hits@1 from 0.379 to 0.549 and Hits@10 from 0.572 to 0.788. Finally, high-shot long-context adaptation further improves unseen-document retrieval to 0.607 Hits@1 and 0.800 Hits@10, with only a modest drop on \(\mathcal{D}_{\text{train}}\). These results show that the \texttt{[COPY]} token enables context retrieval, DPO balances context usage with parametric memory, and long-context adaptation strengthens retrieval under larger candidate sets.
\begin{table}[t]
    \centering
    \caption{Ablation results on MS MARCO with 100-shot in-context demonstrations. Starting from direct inference and the ICL template, we incrementally add the \texttt{[COPY]} token, DPO alignment, and large title context adaptation. Best and second-best results are \textbf{bolded} and \underline{underlined}.}
    \resizebox{\linewidth}{!}{%
    \begin{tabular}{l|cccc}
    \toprule
        \multirow{2}{*}{\textbf{Method}} & \multicolumn{2}{c}{\textbf{Eval on \(\mathcal{D}_{\text{train}}\)}} & \multicolumn{2}{c}{\textbf{Eval on \(\mathcal{D}_{\text{new}}\)}} \\
        \cmidrule(lr){2-3} \cmidrule(lr){4-5}
        & \textbf{Hit@1} & \textbf{Hit@10} & \textbf{Hit@1} & \textbf{Hit@10}  \\
        \midrule
        \multicolumn{3}{l}{\emph{\textbf{Instruct Model}}} \\
         Direct Inference & 0.015 & 0.042 & 0.013 & 0.040 \\
        \midrule
        \multicolumn{3}{l}{\emph{\textbf{Base Model}}} \\
         ICL Template & 0.208 & 0.484 & 0.102 & 0.292 \\
         +\texttt{[COPY]} token & 0.379 & 0.572 & 0.526 & 0.740 \\
         +DPO & \textbf{0.549} & \textbf{0.788} & \underline{0.546} & \underline{0.782} \\
         +Large Context Adap. & \underline{0.519}& \underline{0.772} & \textbf{0.607} & \textbf{0.800} \\
    \bottomrule
    \end{tabular}
    }
    \label{tab:ablation}
\end{table}
\section{Analysis}

\subsection{In-Context Shots Scaling}

\begin{figure}[H]
    \centering
    \includegraphics[width=\columnwidth]{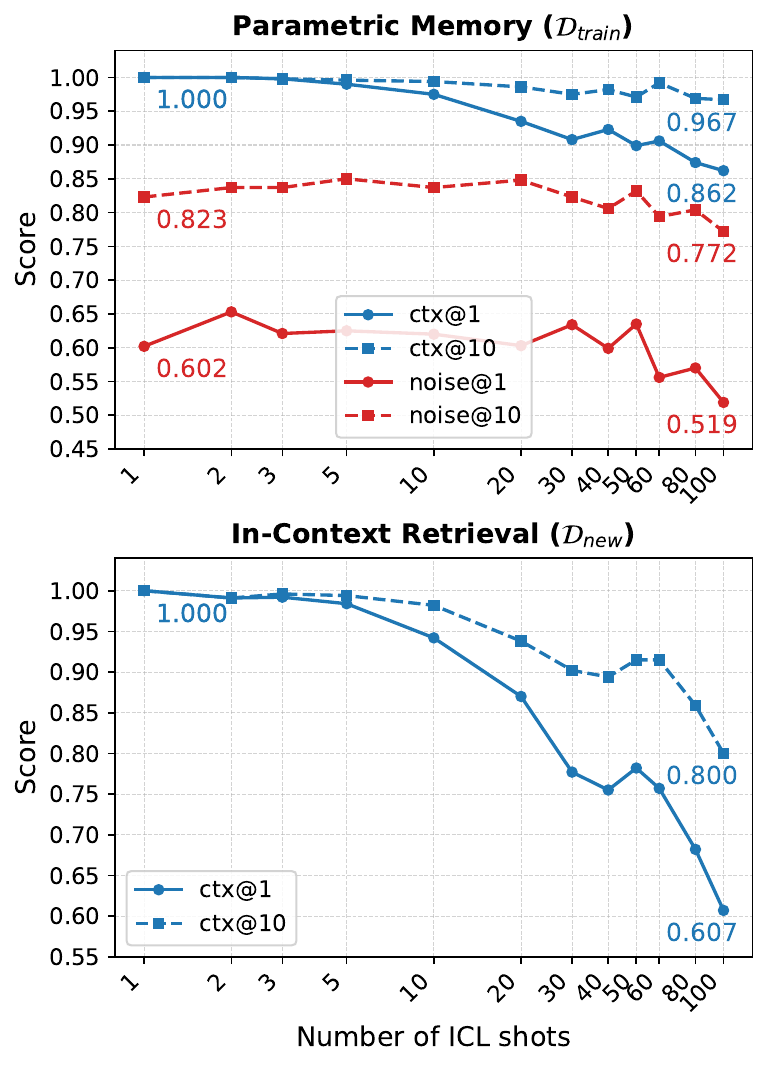}
    \caption{
        Retrieval performance as a function of the number of ICL shots $N$.
        \textbf{Top}: Hits@1 and Hits@10 on seen documents ($\mathcal{D}_{\text{train}}$) under context-dependent (\texttt{ctx}) and query-irrelevant (\texttt{noise}) conditions.
        \textbf{Bottom}: Hits@1 and Hits@10 on unseen documents ($\mathcal{D}_{\text{new}}$) under the context-dependent condition.
        Subscripts @1 and @10 denote Hits@1 and Hits@10 respectively.
    }
    \label{fig:nshot_scaling}
\end{figure}

Figure~\ref{fig:nshot_scaling} shows how retrieval performance scales with the number of in-context shots $N$. For in-context retrieval on unseen documents ($\mathcal{D}_{\text{new}}$, bottom panel), performance is near-perfect when $N\leq10$ but degrades as the candidate pool grows, reaching Hits@1=0.607 and Hits@10=0.800 at $N=100$. A similar trend holds for parametric memory on seen documents ($\mathcal{D}_{\text{train}}$, top panel): both the context-dependent condition (Hits@1: $1.00$ to $0.862$) and the query-irrelevant condition (Hits@1: $0.602$ to $0.519$) degrade with $N$, indicating that longer in-context candidate lists interfere with retrieval in both modes.

However, aggregate Hits@$k$ does not reveal the source of this degradation. In particular, high-shot errors on $\mathcal{D}_{\text{new}}$ may arise from two different failure modes: the model may fail to activate the \texttt{[COPY]} routing signal when the answer is present in the context (\emph{routing failure}), or it may correctly decide to copy but select the wrong candidate document (\emph{retrieval failure}). The fact that ICICLE still achieves Hits@$1$=0.607 at $N=100$ suggests that the copy mechanism does not completely collapse under long contexts. Thus, we further analyze whether errors arise from source selection or from document retrieval after copying in \Cref{subsec:source_selection}.

\subsection{Model Calibration}

\begin{figure}[H]
    \centering
    \includegraphics[width=0.9\linewidth]{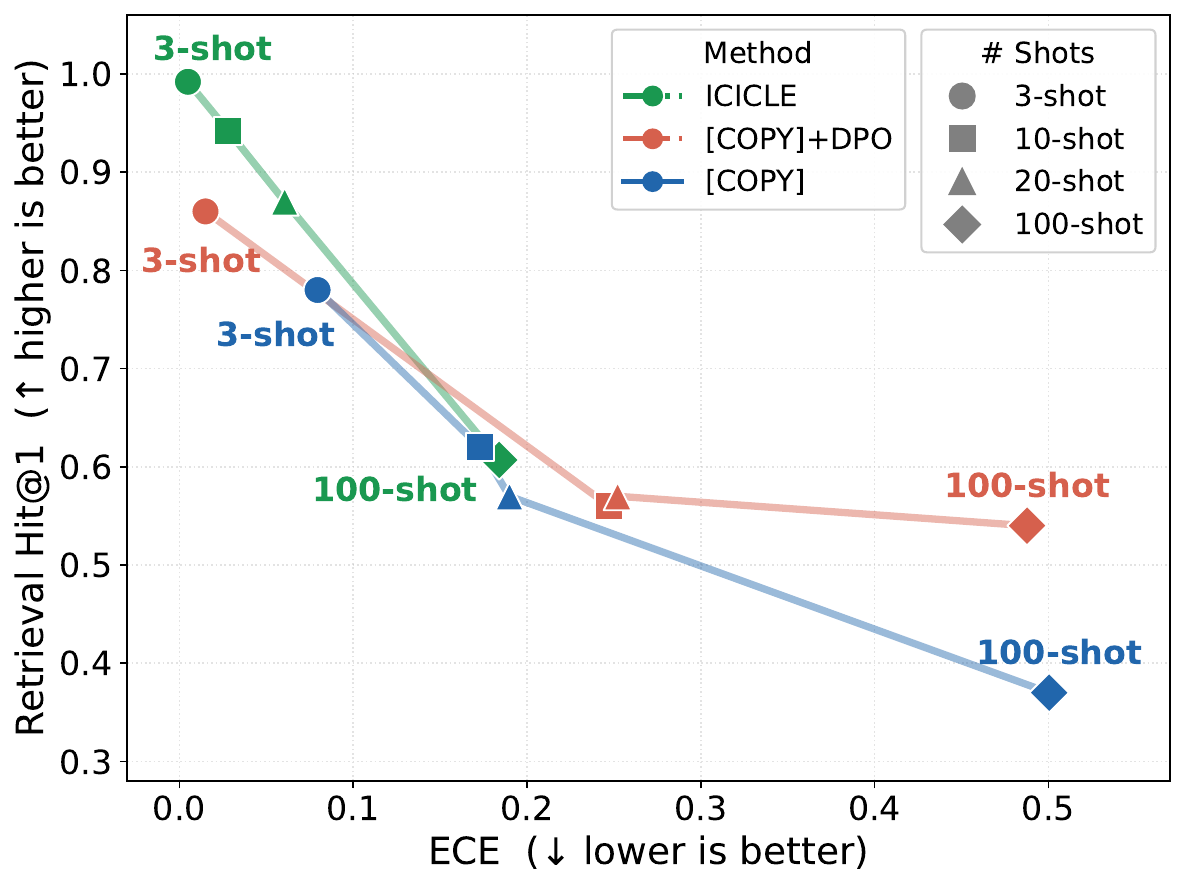}
    \caption{Retrieval Hit@1 vs.\ ECE across methods and shot settings. Each point represents a (method, shot) configuration, lower ECE and higher Hit@1 is better.}
    \label{fig:ece_retrieval}
\end{figure}

We further analyze whether the model can reliably distinguish context-based retrieval from parametric recall. Since the \texttt{[COPY]} token is generated when the model believes that the target document is present in the candidate set, its probability \(s_i = p_\theta(\texttt{[COPY]} \mid q_i, \mathcal{C}_i).\) can be viewed as the model's confidence in using the provided context.
As shown in Figure~\ref{fig:ece_retrieval}, retrieval accuracy and calibration performance are consistently correlated across all methods and shot settings: models with lower calibration error achieve higher retrieval accuracy. Notably, ICICLE maintains low expected calibration error (ECE) as the number of shots increases, while \(\texttt{[COPY]}\) and \(\texttt{[COPY]}\)+DPO become less calibrated under larger candidate pools. This suggests that robust retrieval depends not only on document matching, but also on knowing \emph{when} to use context. Details of ECE are provided in Appendix~\ref{appendix:ece_detail}.

\subsection{Know to Copy or Know to Retrieve?}
\label{subsec:source_selection}

\begin{figure}[H]
    \centering
    \includegraphics[width=\linewidth]{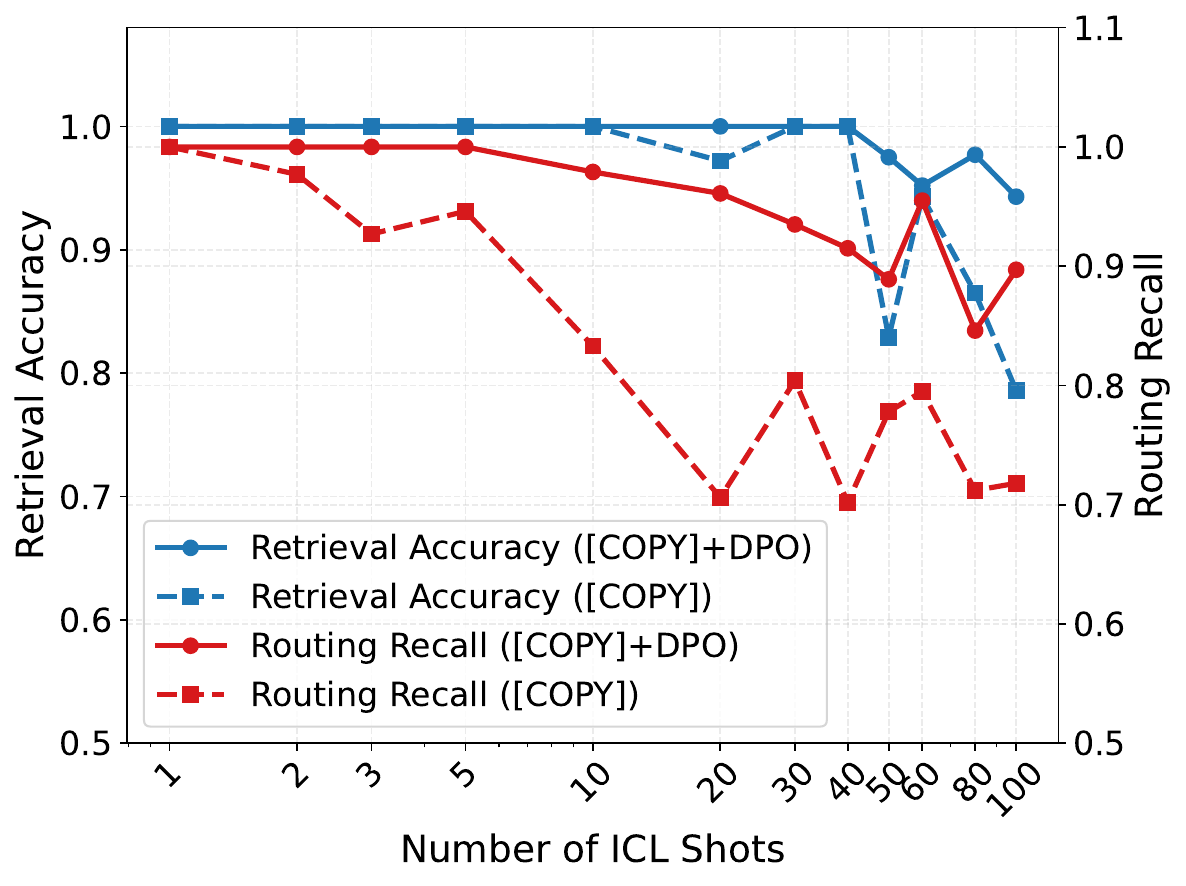}
    \caption{Routing recall and retrieval accuracy as a function of in-context shot count $N$, comparing \(\texttt{[COPY]}\) and \(\texttt{[COPY]}\)+DPO.}
    \label{fig:copy_or_retrieve}
    \vspace{-0.5em}
\end{figure}

To isolate the effect of DPO, we decompose overall retrieval performance into two sub-abilities: \textbf{routing} (correctly generating the \texttt{[COPY]} token when the answer is in context) and \textbf{retrieval} (retrieving the correct answer after deciding to copy). As shown in \Cref{fig:copy_or_retrieve}, with DPO maintains stable conditional retrieval accuracy (\(P(\text{hit} \mid \texttt{[COPY]})\)) at 0.943 to 1.000 across all shot counts, while routing recall monotonically degrades. This indicates that the performance bottleneck under shot scaling lies in routing rather than retrieval. Moreover, the routing-recall gap between the cases of without and with DPO shows that DPO mainly improves source selection under long-context inference. We further shows the source selection analysis on initial corpus \(\mathcal{D}_{\text{train}}\) in Appendix~\ref{appendix:copy_token_routing}.

\section{Related Works}

\paragraph{Generative Retrieval and Incremental Updates}

Generative retrieval (GR) trains a model to map queries directly to document identifiers (docids) via autoregressive decoding~\citep{de2020autoregressive, tay2022transformer, wang2022neural}. Because document-docid associations are encoded into model weights, adding new documents requires parameter updates. Prior work addresses this through incremental training~\citep{mehta2023dsi++, chen2023continual, zhang2025replication, kishore2023incdsi, huynh2025mixlora, son2026cream, guo2025corpusbrain++} or model editing~\citep{zhang2026model}, both of which tie corpus growth to weight modification.  ICICLE instead supplies new document-docid pairs in the input context at inference time, eliminating document-specific parameter updates.

\paragraph{In-Context Learning and Source-Aware Retrieval}

In-context learning (ICL)~\citep{brown2020language, ram2023context, dong2024survey} adapts language models through input demonstrations without weight changes. When applied to GR, it enables retrieval over newly added documents by placing document-docid pairs directly in the prompt. However, this creates a source-selection problem: the model must determine whether to copy a docid from context or recall one from parametric memory. Prior work on knowledge conflicts~\citep{xu2024knowledge} and routing methods such as Self-RAG~\citep{asai2024self}, R-Tuning~\citep{zhang2024r}, and adaptive retrieval~\citep{moskvoretskii2025adaptive} address when to retrieve, but assume fully observable retrieval outputs. ICICLE faces a harder problem: unlike in-context listwise reranking~\citep{gupta2026scalable, tang2024listwise}, which ranks over a fixed observable pool, ICICLE must route between context-provided docids and parametric docids that exist only as implicit memory, with no direct signal indicating what the model has memorized.




\section{Conclusion}

We presented ICICLE, an in-context indexing framework for extending generative retrieval to dynamically expanding corpora without document-specific retraining. ICICLE supplies newly added document-docid associations in the prompt and uses router-aware constrained decoding to decide whether to retrieve from contextual evidence or parametric memory. Experiments on MS MARCO and NQ320K show that ICICLE retrieves unseen corpus at inference time while preserving competitive performance on the original corpus. Our analysis further shows that scaling to larger candidate pools is mainly limited by routing failure rather than ranking difficulty, suggesting calibration as a key factor for reliable in-context generative retrieval.

\section*{Limitations}
While ICICLE provides a practical alternative to full re-indexing for incorporating new documents into a generative retrieval model, its applicability is subject to three boundaries. 
 
First, the number of new documents that ICICLE can accommodate per query is bounded by the model's context window, which was limited by the computation resource on our end. Our current implementation supports at most $100$ in-context document--docid pairs. 
However, 100 is still much smaller than the updates seen in real-world retrieval systems, where thousands of new documents may arrive at once. 
Scaling to this setting would require a two-stage pipeline, in which a coarse pre-retrieval step first reduces the candidate set to a context-feasible size before ICICLE performs the final ranking, and errors at this step would propagate to the final retrieval result. We leave this two-stage extension to future work. 

Second, our shot-scaling analysis show that retrieval accuracy on unseen documents drops as the number of in-context candidates grows. This suggests that ICICLE does suffer from the same context degradation like a LLM \citep{hsiehruler}, however, we believe solutions used in LLM \citep{gao2025train, xiao2024infllm} can be applied to ICICLE as well. 

Third, conditioning on an in-context candidate pool inevitably increases prefill cost relative to query-only generative retrieval baselines, and trie-constrained decoding adds further per-step overhead during generation. Reducing this cost---for example, through adaptive budget allocation of KV-cache across queries \citep{feng2026ada}---is a promising direction for future work.

\FloatBarrier
\renewcommand{\bibname}{References}
\bibliographystyle{plainnat}
\bibliography{custom.bib}

@article{tay2022transformer,
  title={Transformer memory as a differentiable search index},
  author={Tay, Yi and Tran, Vinh and Dehghani, Mostafa and Ni, Jianmo and Bahri, Dara and Mehta, Harsh and Qin, Zhen and Hui, Kai and Zhao, Zhe and Gupta, Jai and others},
  journal={Advances in Neural Information Processing Systems},
  volume={35},
  pages={21831--21843},
  year={2022}
}

@article{de2020autoregressive,
  title={Autoregressive entity retrieval},
  author={De Cao, Nicola and Izacard, Gautier and Riedel, Sebastian and Petroni, Fabio},
  journal={arXiv preprint arXiv:2010.00904},
  year={2020}
}

@article{wang2022neural,
  title={A neural corpus indexer for document retrieval},
  author={Wang, Yujing and Hou, Yingyan and Wang, Haonan and Miao, Ziming and Wu, Shibin and Chen, Qi and Xia, Yuqing and Chi, Chengmin and Zhao, Guoshuai and Liu, Zheng and others},
  journal={Advances in Neural Information Processing Systems},
  volume={35},
  pages={25600--25614},
  year={2022}
}

@inproceedings{mehta2023dsi++,
  title={Dsi++: Updating transformer memory with new documents},
  author={Mehta, Sanket Vaibhav and Gupta, Jai and Tay, Yi and Dehghani, Mostafa and Tran, Vinh Q and Rao, Jinfeng and Najork, Marc and Strubell, Emma and Metzler, Donald},
  booktitle={Proceedings of the 2023 conference on Empirical Methods in Natural Language Processing},
  pages={8198--8213},
  year={2023}
}

@inproceedings{zhang2025replication,
  title={Replication and exploration of generative retrieval over dynamic corpora},
  author={Zhang, Zhen and Ma, Xinyu and Sun, Weiwei and Ren, Pengjie and Chen, Zhumin and Wang, Shuaiqiang and Yin, Dawei and de Rijke, Maarten and Ren, Zhaochun},
  booktitle={Proceedings of the 48th International ACM SIGIR Conference on Research and Development in Information Retrieval},
  pages={3325--3334},
  year={2025}
}

@inproceedings{kishore2023incdsi,
  title={Incdsi: Incrementally updatable document retrieval},
  author={Kishore, Varsha and Wan, Chao and Lovelace, Justin and Artzi, Yoav and Weinberger, Kilian Q},
  booktitle={International Conference on Machine Learning},
  pages={17122--17134},
  year={2023},
  organization={PMLR}
}

@inproceedings{zhang2026model,
  title={Model Editing for New Document Integration in Generative Information Retrieval},
  author={Zhang, Zhen and Wang, Zihan and Ma, Xinyu and Wang, Shuaiqiang and Yin, Dawei and Xin, Xin and Ren, Pengjie and de Rijke, Maarten and Ren, Zhaochun},
  booktitle={Proceedings of the ACM Web Conference 2026},
  pages={1993--2003},
  year={2026}
}

@article{brown2020language,
  title={Language models are few-shot learners},
  author={Brown, Tom and Mann, Benjamin and Ryder, Nick and Subbiah, Melanie and Kaplan, Jared D and Dhariwal, Prafulla and Neelakantan, Arvind and Shyam, Pranav and Sastry, Girish and Askell, Amanda and others},
  journal={Advances in Neural Information Processing Systems},
  volume={33},
  pages={1877--1901},
  year={2020}
}

@inproceedings{lee2023nonparametric,
  title={Nonparametric decoding for generative retrieval},
  author={Lee, Hyunji and Kim, Jaeyoung and Chang, Hoyeon and Oh, Hanseok and Yang, Sohee and Karpukhin, Vladimir and Lu, Yi and Seo, Minjoon},
  booktitle={Findings of the Association for Computational Linguistics: ACL 2023},
  pages={12642--12661},
  year={2023}
}

@article{NguyenRSGTMD16,
  author    = {Tri Nguyen and
               Mir Rosenberg and
               Xia Song and
               Jianfeng Gao and
               Saurabh Tiwary and
               Rangan Majumder and
               Li Deng},
  title     = {{MS} {MARCO:} {A} Human Generated MAchine Reading COmprehension Dataset},
  journal   = {CoRR},
  volume    = {abs/1611.09268},
  year      = {2016},
}

@article{47761,
title	= {Natural Questions: a Benchmark for Question Answering Research},
author	= {Tom Kwiatkowski and Jennimaria Palomaki and Olivia Redfield and Michael Collins and Ankur Parikh and Chris Alberti and Danielle Epstein and Illia Polosukhin and Matthew Kelcey and Jacob Devlin and Kenton Lee and Kristina N. Toutanova and Llion Jones and Ming-Wei Chang and Andrew Dai and Jakob Uszkoreit and Quoc Le and Slav Petrov},
year	= {2019},
journal	= {Transactions of the Association of Computational Linguistics}
}

@article{tang2024listwise,
  title={Listwise generative retrieval models via a sequential learning process},
  author={Tang, Yubao and Zhang, Ruqing and Guo, Jiafeng and De Rijke, Maarten and Chen, Wei and Cheng, Xueqi},
  journal={ACM Transactions on Information Systems},
  volume={42},
  number={5},
  pages={1--31},
  year={2024},
  publisher={ACM New York, NY}
}

@inproceedings{chen2023continual,
  title={Continual learning for generative retrieval over dynamic corpora},
  author={Chen, Jiangui and Zhang, Ruqing and Guo, Jiafeng and de Rijke, Maarten and Chen, Wei and Fan, Yixing and Cheng, Xueqi},
  booktitle={Proceedings of the 32nd ACM International Conference on Information and Knowledge Management},
  pages={306--315},
  year={2023}
}

@book{robertson2009probabilistic,
  title={The probabilistic relevance framework: BM25 and beyond},
  author={Robertson, Stephen and Zaragoza, Hugo},
  volume={4},
  year={2009},
  publisher={Now Publishers Inc}
}

@inproceedings{karpukhin2020dense,
  title={Dense passage retrieval for open-domain question answering},
  author={Karpukhin, Vladimir and Oguz, Barlas and Min, Sewon and Lewis, Patrick and Wu, Ledell and Edunov, Sergey and Chen, Danqi and Yih, Wen-tau},
  booktitle={Proceedings of the 2020 conference on empirical methods in natural language processing (EMNLP)},
  pages={6769--6781},
  year={2020}
}

@inproceedings{dong2024survey,
  title = {A Survey on In-context Learning},
  author = {Dong, Qingxiu and Li, Lei and Dai, Damai and Zheng, Ce and Ma, Jingyuan and Li, Rui and Xia, Heming and Xu, Jingjing and Wu, Zhiyong and Chang, Baobao and others},
  booktitle = {Proceedings of the 2024 Conference on Empirical Methods in Natural Language Processing},
  year = {2024}
}

@inproceedings{min2022rethinking,
  title = {Rethinking the Role of Demonstrations: What Makes In-Context Learning Work?},
  author = {Min, Sewon and Lyu, Xinxi and Holtzman, Ari and Artetxe, Mikel and Lewis, Mike and Hajishirzi, Hannaneh and Zettlemoyer, Luke},
  booktitle = {Proceedings of the 2022 Conference on Empirical Methods in Natural Language Processing},
  year = {2022}
}

@article{rafailov2023direct,
  title={Direct preference optimization: Your language model is secretly a reward model},
  author={Rafailov, Rafael and Sharma, Archit and Mitchell, Eric and Manning, Christopher D and Ermon, Stefano and Finn, Chelsea},
  journal={Advances in neural information processing systems},
  volume={36},
  pages={53728--53741},
  year={2023}
}

@inproceedings{chen2024m3,
  title={M3-embedding: Multi-linguality, multi-functionality, multi-granularity text embeddings through self-knowledge distillation},
  author={Chen, Jianlyu and Xiao, Shitao and Zhang, Peitian and Luo, Kun and Lian, Defu and Liu, Zheng},
  booktitle={Findings of the association for computational linguistics: ACL 2024},
  pages={2318--2335},
  year={2024}
}

@article{sun2023learning,
  title={Learning to tokenize for generative retrieval},
  author={Sun, Weiwei and Yan, Lingyong and Chen, Zheng and Wang, Shuaiqiang and Zhu, Haichao and Ren, Pengjie and Chen, Zhumin and Yin, Dawei and Rijke, Maarten and Ren, Zhaochun},
  journal={Advances in Neural Information Processing Systems},
  volume={36},
  pages={46345--46361},
  year={2023}
}

@inproceedings{huynh2025mixlora,
  title={MixLoRA-DSI: Dynamically Expandable Mixture-of-LoRA Experts for Rehearsal-Free Generative Retrieval over Dynamic Corpora},
  author={Huynh, Tuan-Luc and Vu, Thuy and Wang, Weiqing and Le, Trung and Gasevic, Dragan and Li, Yuan-Fang and Do, Thanh-Toan},
  booktitle={Proceedings of the 2025 Conference on Empirical Methods in Natural Language Processing},
  pages={380--396},
  year={2025}
}

@article{gupta2026scalable,
  title={Scalable in-context ranking with generative models},
  author={Gupta, Nilesh and You, Chong and Bhojanapalli, Srinadh and Kumar, Sanjiv and Dhillon, Inderjit and Yu, Felix},
  journal={Advances in Neural Information Processing Systems},
  volume={38},
  pages={36590--36613},
  year={2026}
}

@article{ram2023context,
  title={In-context retrieval-augmented language models},
  author={Ram, Ori and Levine, Yoav and Dalmedigos, Itay and Muhlgay, Dor and Shashua, Amnon and Leyton-Brown, Kevin and Shoham, Yoav},
  journal={Transactions of the Association for Computational Linguistics},
  volume={11},
  pages={1316--1331},
  year={2023},
  publisher={MIT Press One Broadway, 12th Floor, Cambridge, Massachusetts 02142, USA~…}
}

@inproceedings{lee2023glen,
  title={GLEN: Generative retrieval via lexical index learning},
  author={Lee, Sunkyung and Choi, Minjin and Lee, Jongwuk},
  booktitle={Proceedings of the 2023 Conference on Empirical Methods in Natural Language Processing},
  pages={7693--7704},
  year={2023}
}

@inproceedings{shi2024trusting,
  title={Trusting your evidence: Hallucinate less with context-aware decoding},
  author={Shi, Weijia and Han, Xiaochuang and Lewis, Mike and Tsvetkov, Yulia and Zettlemoyer, Luke and Yih, Wen-tau},
  booktitle={Proceedings of the 2024 Conference of the North American Chapter of the Association for Computational Linguistics: Human Language Technologies (Volume 2: Short Papers)},
  pages={783--791},
  year={2024}
}

@inproceedings{son2026cream,
  title={CREAM: Continual Retrieval on Dynamic Streaming Corpora with Adaptive Soft Memory},
  author={Son, HuiJeong and Kang, Hyeongu and Kim, Sunho and Ho, Subeen and Kang, SeongKu and Lee, Dongha and Yoon, Susik},
  booktitle={Proceedings of the 32nd ACM SIGKDD Conference on Knowledge Discovery and Data Mining V. 1},
  pages={1297--1308},
  year={2026}
}

@inproceedings{lawrie2024plaid,
  title={PLAID SHIRTTT for large-scale streaming dense retrieval},
  author={Lawrie, Dawn and Kayi, Efsun and Yang, Eugene and Mayfield, James and Oard, Douglas W},
  booktitle={Proceedings of the 47th International ACM SIGIR Conference on Research and Development in Information Retrieval},
  pages={2574--2578},
  year={2024}
}

@article{guo2025corpusbrain++,
  title={Corpusbrain++: A continual generative pre-training framework for knowledge-intensive language tasks},
  author={Guo, Jiafeng and Zhou, Changjiang and Zhang, Ruqing and Chen, Jiangui and de Rijke, Maarten and Fan, Yixing and Cheng, Xueqi},
  journal={ACM Transactions on Information Systems},
  volume={44},
  number={1},
  pages={1--35},
  year={2025},
  publisher={ACM New York, NY}
}

@inproceedings{zhang2024r,
  title={R-tuning: Instructing large language models to say ‘i don’t know’},
  author={Zhang, Hanning and Diao, Shizhe and Lin, Yong and Fung, Yi and Lian, Qing and Wang, Xingyao and Chen, Yangyi and Ji, Heng and Zhang, Tong},
  booktitle={Proceedings of the 2024 Conference of the North American Chapter of the Association for Computational Linguistics: Human Language Technologies (Volume 1: Long Papers)},
  pages={7113--7139},
  year={2024}
}

@inproceedings{moskvoretskii2025adaptive,
  title={Adaptive retrieval without self-knowledge? bringing uncertainty back home},
  author={Moskvoretskii, Viktor and Marina, Maria and Salnikov, Mikhail and Ivanov, Nikolay and Pletenev, Sergey and Galimzianova, Daria and Krayko, Nikita and Konovalov, Vasily and Nikishina, Irina and Panchenko, Alexander},
  booktitle={Proceedings of the 63rd Annual Meeting of the Association for Computational Linguistics (Volume 1: Long Papers)},
  pages={6355--6384},
  year={2025}
}

@inproceedings{asai2024self,
  title={Self-rag: Learning to retrieve, generate, and critique through self-reflection},
  author={Asai, Akari and Wu, Zeqiu and Wang, Yizhong and Sil, Avi and Hajishirzi, Hannaneh},
  booktitle={International conference on learning representations},
  volume={2024},
  pages={9112--9141},
  year={2024}
}

@article{yang2023longqlora,
  title={Longqlora: Efficient and effective method to extend context length of large language models},
  author={Yang, Jianxin},
  journal={arXiv preprint arXiv:2311.04879},
  year={2023}
}

@article{mekonnen2026parametric,
  title={A Parametric Memory Head for Continual Generative Retrieval},
  author={Mekonnen, Kidist Amde and Tang, Yubao and de Rijke, Maarten},
  journal={arXiv preprint arXiv:2604.23388},
  year={2026}
}

@article{yang2025qwen3,
  title={Qwen3 technical report},
  author={Yang, An and Li, Anfeng and Yang, Baosong and Zhang, Beichen and Hui, Binyuan and Zheng, Bo and Yu, Bowen and Gao, Chang and Huang, Chengen and Lv, Chenxu and others},
  journal={arXiv preprint arXiv:2505.09388},
  year={2025}
}

@article{rajput2023recommender,
  title={Recommender systems with generative retrieval},
  author={Rajput, Shashank and Mehta, Nikhil and Singh, Anima and Hulikal Keshavan, Raghunandan and Vu, Trung and Heldt, Lukasz and Hong, Lichan and Tay, Yi and Tran, Vinh and Samost, Jonah and others},
  journal={Advances in Neural Information Processing Systems},
  volume={36},
  pages={10299--10315},
  year={2023}
}

@article{bge_embedding,
      title={C-Pack: Packaged Resources To Advance General Chinese Embedding}, 
      author={Shitao Xiao and Zheng Liu and Peitian Zhang and Niklas Muennighoff},
      year={2023},
      journal={arXiv preprint arXiv:2309.07597}
}

@inproceedings{xu2024knowledge,
  title={Knowledge conflicts for llms: A survey},
  author={Xu, Rongwu and Qi, Zehan and Guo, Zhijiang and Wang, Cunxiang and Wang, Hongru and Zhang, Yue and Xu, Wei},
  booktitle={Proceedings of the 2024 Conference on Empirical Methods in Natural Language Processing},
  pages={8541--8565},
  year={2024}
}

@inproceedings{hsiehruler,
  title={RULER: What’s the Real Context Size of Your Long-Context Language Models?},
  author={Hsieh, Cheng-Ping and Sun, Simeng and Kriman, Samuel and Acharya, Shantanu and Rekesh, Dima and Jia, Fei and Ginsburg, Boris},
  booktitle={First Conference on Language Modeling},
  year={2024}
}

@article{feng2026ada,
  title={Ada-kv: Optimizing kv cache eviction by adaptive budget allocation for efficient llm inference},
  author={Feng, Yuan and Lv, Junlin and Cao, Yukun and Xie, Xike and Zhou, S Kevin},
  journal={Advances in Neural Information Processing Systems},
  volume={38},
  pages={113152--113188},
  year={2026}
}

@inproceedings{gao2025train,
  title={How to train long-context language models (effectively)},
  author={Gao, Tianyu and Wettig, Alexander and Yen, Howard and Chen, Danqi},
  booktitle={Proceedings of the 63rd Annual Meeting of the Association for Computational Linguistics (Volume 1: Long Papers)},
  pages={7376--7399},
  year={2025}
}

@article{xiao2024infllm,
  title={Infllm: Training-free long-context extrapolation for llms with an efficient context memory},
  author={Xiao, Chaojun and Zhang, Pengle and Han, Xu and Xiao, Guangxuan and Lin, Yankai and Zhang, Zhengyan and Liu, Zhiyuan and Sun, Maosong},
  journal={Advances in neural information processing systems},
  volume={37},
  pages={119638--119661},
  year={2024}
}

\newpage
\onecolumn

\appendix 
\appendix
\onecolumn
\section{In-Context Template Learning}

\subsection{In-Context Template}
\label{appendix:icl_template}
We use a unified in-context template to support two retrieval modes within the same GR model. Given a query and a candidate set, the model must either copy the docid of a relevant candidate from context or fall back to its parametric memory when the relevant document is absent from the context. The special token \texttt{[COPY]} serves as an explicit routing signal for the former case. When the gold document appears in the candidate list, the target output is \texttt{[COPY]} followed by the corresponding docid. When the context contains only irrelevant documents, the target output is the gold docid without \texttt{[COPY]}, forcing the model to retrieve from its learned global document-docid mapping.

\tcbset{boxrule = 0.5mm, colback=blue!5!white, colframe=blue!30!black,
  title = {In-Context Template}}
\begin{tcolorbox}
\textbf{System:} Given a query and a list of candidate documents, retrieve the title of the most relevant document. If the relevant document appears in the candidate list, output \texttt{[COPY]} followed by its title. Otherwise, output the title from memory directly.

\vspace{0.5em}
\textbf{Candidates:}

\textbf{[1]} \textit{\{compressed content of document 1\}}\\
\hspace*{1em} Title: \textit{\{title 1\}}

\textbf{[2]} \textit{\{compressed content of document 2\}}\\
\hspace*{1em} Title: \textit{\{title 2\}}

\hspace*{1.2em} $\vdots$

\textbf{[n]} \textit{\{compressed content of document n\}}\\
\hspace*{1em} Title: \textit{\{title n\}}

\vspace{0.5em}
\textbf{Query:} \textit{\{query\}}

\vspace{0.3em}
\textbf{Output (context-dependent):} \texttt{[COPY]} \textit{\{gold title\}}\\
\textbf{Output (all-noise):} \textit{\{gold title\}}
\end{tcolorbox}

\subsection{Details of Hard Negatives Mining}
\label{appendix:hard_neg}

For each training document $d \in \mathcal{D}_{\text{train}}$, we construct a hard negative set $\mathcal{H}(d)$ by retrieving its nearest semantic neighbors using a dense retrieval model. Concretely, we encode every documents in $\mathcal{D}_{\text{train}}$ with \texttt{bge-base-en-v1.5}~\citep{bge_embedding} and $\ell_2$-normalize the resulting embedding.

We then index the pool with an exact inner-product (FAISS \texttt{IndexFlatIP}). For each document $d$, we retrieve its top-$k$ most similar documents, excluding $d$ itself, and take them as $\mathcal{H}(d)$; we set $k = $ 100 in all experiments.

When constructing in-context instances, the irrelevant slots in $\mathcal{C}_i$ are filled by sampling without replacement from $\mathcal{H}(d_i^\star)$ rather than from randomly drawn documents in $\mathcal{D}_{\text{train}}$. Because hard negatives share topical content with $d_i^\star$, the model cannot rely on superficial topical differences to identify the correct candidate and must instead learn fine-grained semantic distinctions and faithfully use the provided document--DocID mapping.

\section{Routing Between Contextual and Parametric Retrieval}
\subsection{Details of Context-Aware Constrained Decoding}
\label{appendix:constrained_decoding}
We modify the trie-based constrained decoding to account for the \texttt{[COPY]} routing token. At the first decoding step, the model must choose between generating \texttt{[COPY]} or directly generating a valid global docid prefix. This first token determines the decoding route: if \texttt{[COPY]} is generated, the remaining tokens are constrained by the context trie \(\mathcal{T}_i^C\), ensuring that the prediction is copied from the candidate set; otherwise, decoding is constrained by the global trie \(\mathcal{T}_G\), corresponding to parametric retrieval from the full corpus. Thus, the decoding procedure preserves valid docid generation while enforcing the intended separation between context-dependent and query-irrelevant retrieval.

\begin{algorithm}[H]
  \caption{Context-Aware Constrained Decoding}
  \label{alg:copy_constrained_decoding}
  \begin{algorithmic}[1]

  \Require Global trie $\mathcal{T}_G$, context trie $\mathcal{T}_i^C$ for sample $i$
  \Require Generated tokens $y_{1:t}$ (excluding prompt)
  \Require Logits $\mathbf{s} \in \mathbb{R}^{|\mathcal{V}|}$

  \Statex

  \If{$t = 0$} \Comment{force routing decision}
      \State allow only $[\texttt{COPY}]$ or valid starts of $\mathcal{T}_G$
  \ElsIf{$y_1 = [\texttt{COPY}]$}
      \State constrain $\mathbf{s}$ using $\mathcal{T}_i^C$ on $y_{2:t}$ \Comment{copy from context}
  \Else
      \State constrain $\mathbf{s}$ using $\mathcal{T}_G$ on $y_{1:t}$ \Comment{retrieve from corpus}
  \EndIf
  \end{algorithmic}
\end{algorithm}

\subsection{Details of DPO}
\label{appendix:dpo}
Let \(\pi_{\theta}\) denote the policy model and \(\pi_{\mathrm{ref}}\) denote the frozen reference model initialized from the SFT checkpoint. 
We optimize the DPO objective:

\begin{equation}
    \mathcal{L}_{\mathrm{DPO}}
    =
    -\mathbb{E}_{(\mathcal{C},q,y^{+},y^{-})}
    \log \sigma
    \left(
    \beta
    \left[
    \log \frac{\pi_{\theta}(y^{+}\mid \mathcal{C},q)}
              {\pi_{\mathrm{ref}}(y^{+}\mid \mathcal{C},q)}
    -
    \log \frac{\pi_{\theta}(y^{-}\mid \mathcal{C},q)}
              {\pi_{\mathrm{ref}}(y^{-}\mid \mathcal{C},q)}
    \right]
    \right).
\end{equation}
This objective explicitly encourages the model to prefer context-grounded copying over context-blind parametric fallback when the answer is available in the prompt, while also improving the ranking of correct identifiers among beam candidates.

\section{Details of Large Title Context Adaptation}
\label{app:context_adaptation}

The DPO stage is optimized with a 3-shot template, but inference uses up to $K=100$ in-context candidates. To bridge this gap, we perform a LoRA-based adaptation stage after DPO, using an identifier-only context $\tilde{\mathcal{C}}_K = \{y_i\}_{i=1}^{K}$ in which only document titles (identifiers) appear, with no document content. This reduces the per-candidate footprint to a short token sequence and allows 100 candidates to fit within the model's context window.

\paragraph{Training data construction.} For each training query $q_i$ with gold docid $y_i^\star$, we sample $K-1$ distractor identifiers from $\mathcal{D}_{\text{train}}$ and insert $y_i^\star$ at a uniformly sampled position, forming $\tilde{\mathcal{C}}_K$. We include both context-dependent instances (gold present in $\tilde{\mathcal{C}}_K$) and all-noise instances (gold absent, distractors only). The training objective is identical to $\mathscr{L}_{\text{ICL}}$ in \Cref{sec:method}, applied to the identifier-only context.

\section{Details of Datasets}
\label{sec:data}

\paragraph{MS MARCO.} We use the MS MARCO passage retrieval collection~\citep{NguyenRSGTMD16}, which contains near 100K passages. We adopt the standard training queries and randomly partition the corpus into 90\% ($\mathcal{D}_{\text{train}}$, 90K passages) used to train the base GR model and 10\% ($\mathcal{D}_{\text{new}}$, 10K passages) withheld as the unseen document set.

\paragraph{NQ320K.} We use the Natural Questions~\citep{47761} retrieval benchmark with also near 100K passages. The same 90/10 corpus partition is applied, yielding $\mathcal{D}_{\text{train}}$ (90K passages) and $\mathcal{D}_{\text{new}}$ (10K passages).

For both datasets, test queries are split according to whether their relevant document belongs to $\mathcal{D}_{\text{train}}$ or $\mathcal{D}_{\text{new}}$, and performance is reported separately for each subset.

\section{Details of Baselines}
\label{appendix:baselines}
For sparse retrieval, we implement BM25 using the \texttt{bm25s} package\footnote{\url{https://github.com/xhluca/bm25s}}. For dense retrieval, we build DPR with \texttt{Pyserini}\footnote{\url{https://github.com/castorini/pyserini}}. We use our own implementations for DSI, DSI++, and the two oracle-style incremental learning baselines, \textsc{From-Scratch} and \textsc{New Doc FT}. The two oracle-style baselines share the same hyperparameter configuration reported below, with the only difference being the number of training epochs: \textsc{From-Scratch} is trained for 15 epochs, whereas \textsc{New Doc FT} is trained for 10 epochs. For DOME, we directly use the authors' official implementation\footnote{\url{https://github.com/zhangzhen-research/DOME}}.

\begin{table}[h]
    \centering
    \label{tab:hyperparams}
    \begin{tabular}{lc}
    \toprule
    \textbf{Hyperparameter} & \textbf{Value} \\
    \midrule
    Backbone & Qwen3-1.7B-Base \\
    learning rate & $1\times10^{-4}$ \\
    Warmup ratio & 0.1 \\
    training epochs & 10/15 \\
    Effective batch size & 128 \\
    Optimizer & AdamW \\
    Maximum sequence length $L$ & 2048 \\
    \bottomrule
    \end{tabular}
\end{table}

\section{ICICLE Training Configurations}
\label{appendix:training_conf}

We perform the training of the ICICLE model entirely on RTX PRO 6000 Blackwell.

\subsection{Stage-1 SFT Training}

Stage-1 SFT consists of two consecutive supervised finetuning phases. First, we train the model with indexing conversations to establish the document--DocID mapping ability required by GR. Second, starting from this indexing checkpoint, we continue SFT on in-context instances constructed as described in \Cref{subsec:icl_template}. For each training query, we generate one context-dependent instance and one all-noise instance, so that each optimization step sees both retrieval modes. Hard negatives are used as irrelevant documents in both instance types (Appendix~\ref{appendix:hard_neg}). We further mix indexing-style conversations into this phase as anchor data, which preserves the learned identifier mappings while the model adapts to in-context retrieval.

\begin{table}[h]
    \centering
    \label{tab:hyperparams}
    \begin{tabular}{lc}
    \toprule
    \textbf{Hyperparameter} & \textbf{Value} \\
    \midrule
    Backbone & Qwen3-1.7B-Base \\
    Indexing SFT learning rate & $1\times10^{-4}$ \\
    Indexing SFT training epochs & 10 \\
    ICL SFT learning rate & $5\times10^{-5}$ \\
    ICL SFT effective batch size & 512 \\
    Shots $n$ per instance & 3 \\
    Maximum sequence length $L$ & 4096 \\
    Warmup ratio & 0.1 \\
    Optimizer & AdamW \\
    Special token & \texttt{[COPY]} \\
    \bottomrule
    \end{tabular}
\end{table}

\subsection{Stage-2 DPO Training}

DPO preference pairs are constructed as described in \Cref{sec:method}, covering both routing failures and ranking failures. We use a 3-shot template for all DPO instances.

\begin{table}[h]
    \centering
    \label{tab:dpo_hyperparams}
    \begin{tabular}{lc}
    \toprule
    \textbf{Hyperparameter} & \textbf{Value} \\
    \midrule
    $\beta$ & 0.1 \\
    Learning rate & $5\times10^{-6}$ \\
    Effective batch size & 64 \\
    Training epochs & 3 \\
    Warmup ratio & 0.1 \\
    Optimizer & AdamW \\
    \midrule
    \multicolumn{2}{l}{\textbf{\textit{Number of preference pairs (routing / ranking)}}} \\
    \quad NQ320K & 8435 / 8435 \\
    \quad MS MARCO & 9160 / 9160 \\
    \bottomrule
    \end{tabular}
\end{table}

\subsection{Large Title Context Adaptation}

After DPO, we perform the large title context adaptation stage described. This stage uses title-only contexts with up to $K=100$ candidates and trains LoRA adapters while keeping the backbone frozen. The purpose is to reduce the mismatch between the short 3-shot training template and the long candidate lists used at inference time. Since only identifiers are included in the context, the adaptation stage is computationally lighter than full document-based long-context training and allows the model to observe many more candidate positions.

\begin{table}[h]
    \centering
    \label{tab:lora_hyperparams}
    \begin{tabular}{lc}
    \toprule
    \textbf{Hyperparameter} & \textbf{Value} \\
    \midrule
    LoRA rank $r$ & 16 \\
    LoRA $\alpha$ & 16 \\
    Dropout rate & 0.05 \\
    Learning rate & $1\times10^{-5}$ \\
    Warmup ratio & 0.1 \\
    Optimizer & AdamW \\
    Training epochs & 2 \\
    Frozen parameters & Backbone $\theta$ \\
    \midrule
    \multicolumn{2}{l}{\textbf{\textit{Target modules}}} \\
    \quad Attention & \texttt{q\_proj}, \texttt{k\_proj}, \texttt{v\_proj}, \texttt{o\_proj} \\
    \quad Feed-forward & \texttt{gate\_proj}, \texttt{up\_proj}, \texttt{down\_proj} \\
    \bottomrule
    \end{tabular}
\end{table}

\section{Full Experiment Results over $N$-shots}

Table~\ref{tab:full_shot} reports Hits@1 on both $\mathcal{D}_{\text{train}}$ and $\mathcal{D}_{\text{new}}$ for each in-context shot count $N \in \{3, 10, 20, 50, 100\}$ under the context-dependent (\texttt{ctx}) and all-noise (\texttt{noise}) conditions.

\begin{table}[h]
    \centering
    \caption{$N$-shot result of \(\mathcal{D}_{\text{train}}\)}
    \begin{tabular}{lcccccccccc}
    \toprule
    & \multicolumn{5}{c}{\textbf{MS MARCO}} & \multicolumn{5}{c}{\textbf{NQ320K}} \\
    \cmidrule(lr){2-6} \cmidrule(lr){7-11}
    \textit{\# shots} $\rightarrow$ & 3 & 10 & 20 & 50 & 100 & 3 & 10 & 20 & 50 & 100 \\
    \midrule
    \texttt{[COPY]} & 0.503 & 0.508 & 0.518 & 0.448 & 0.379 & 0.280 & 0.300 & 0.300 & 0.320 & 0.356 \\
    \texttt{[COPY]} + DPO & 0.602 & 0.610 & 0.609 & 0.630 & 0.549 & 0.300 & 0.432 & 0.448 & 0.524 & 0.524 \\
    ICICLE & 0.621 & 0.620 & 0.603 & 0.635 & 0.519 & 0.236 & 0.396 & 0.400 & 0.500 & 0.483\\
    \bottomrule
    \end{tabular}
    \label{tab:full_shot}
\end{table}

\begin{table}[h]
    \centering
    \caption{$N$-shot result of \(\mathcal{D}_{\text{new}}\)}
    \begin{tabular}{lcccccccccc}
    \toprule
    & \multicolumn{5}{c}{\textbf{MS MARCO}} & \multicolumn{5}{c}{\textbf{NQ320K}} \\
    \cmidrule(lr){2-6} \cmidrule(lr){7-11}
    \textit{\# shots} $\rightarrow$ & 3 & 10 & 20 & 50 & 100 & 3 & 10 & 20 & 50 & 100 \\
    \midrule
    \texttt{[COPY]} & 0.959 & 0.868 & 0.737 & 0.719 & 0.526 & 0.600 & 0.712 & 0.664 & 0.592 & 0.416 \\
    \texttt{[COPY]} + DPO & 0.984 & 0.914 & 0.715 & 0.624 & 0.546 & 0.804 & 0.728 & 0.664 & 0.568 & 0.556 \\
    ICICLE & 0.992 & 0.942 & 0.870 & 0.782 & 0.607 & 0.832 & 0.780 & 0.732 & 0.624 & 0.649 \\
    \bottomrule
    \end{tabular}
    \label{tab:full_shot}
\end{table}

\section{Details of ECE}
\label{appendix:ece_detail}

For each instance, we define the ground-truth mode as
\begin{equation}
    z_i =
    \begin{cases}
    1, & \text{if } d_i^\star \in \mathcal{C}_i,\\
    0, & \text{otherwise},
    \end{cases}
\end{equation}

\noindent where \(d_i^*\) is the target document and \(\mathcal{C}_i\) is the context. We use the \texttt{[COPY]} probability $s_i$ as the confidence score for context-based retrieval. ECE measures the discrepancy between the model's confidence and empirical correctness across confidence bins:
\begin{equation}
    \mathrm{ECE}
    =
    \sum_{m=1}^{M}
    \frac{|B_m|}{N}
    \left|
    \mathrm{acc}(B_m) - \mathrm{conf}(B_m)
    \right|,
\end{equation}

\noindent where

\begin{equation}
    \mathrm{acc}(B_m)
    =
    \frac{1}{|B_m|}
    \sum_{i \in B_m}
    \mathbb{I}[\hat{z}_i = z_i],
\end{equation}
and
\begin{equation}
    \mathrm{conf}(B_m)
    =
    \frac{1}{|B_m|}
    \sum_{i \in B_m}
    \max(s_i, 1-s_i).
\end{equation}

A lower ECE indicates that the model's confidence better reflects the actual correctness of its retrieval-mode decision.



\section{\texttt{[COPY]} Token Routing Probability}
\label{appendix:copy_token_routing}

\begin{figure}[h]
    \centering
    \includegraphics[width=\linewidth]{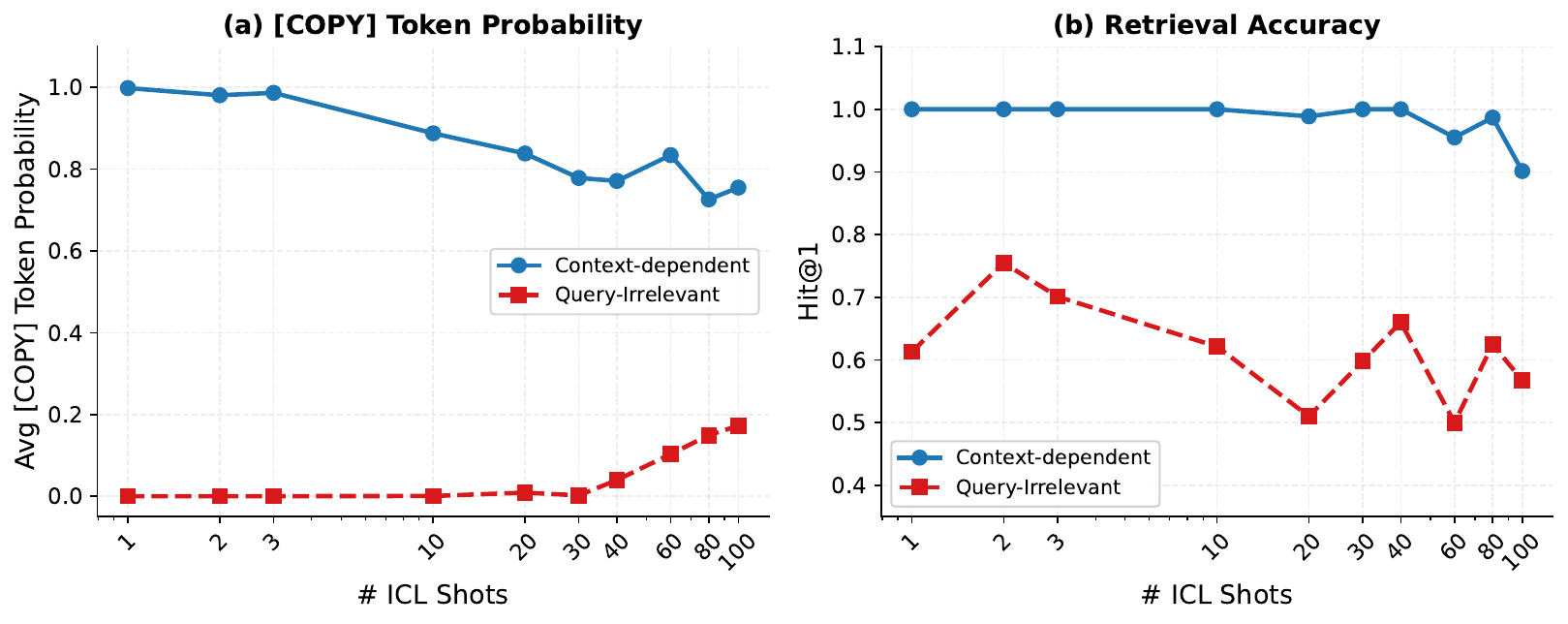}
    \caption{[COPY] token probability and retrieval accuracy under two context conditions on the same test queries. Context-dependent: the answer is present in context. All-noise: context contains only irrelevant documents. All test documents were seen during training.}
    \label{fig:copy_prob}
\end{figure}

We analyze how the model decides between context retrieval and parametric memory. This analysis is conducted on the test set, where each query is evaluated under two conditions: a context-dependent condition in which the answer is provided in the ICL context, and an all-noise condition in which the same query must be answered from parametric memory. As shown in \Cref{fig:copy_prob}(a), when the answer is present in context, the model assigns near-unity probability to the \texttt{[COPY]} token at low shot counts, despite being capable of retrieving the same answers from parametric memory---evidenced by the \(\sim\)0.65 Hit@1 achieved under the all-noise condition on identical queries (\Cref{fig:copy_prob}(b)). This suggests that the model treats a short, focused context as a stronger evidence signal than its own encoded knowledge. However, this preference weakens monotonically as the number of ICL shots increases, and retrieval accuracy degrades accordingly. We further show the scenario of \texttt{[COPY]} token generation error in below table.

\begin{table}[h]
  \centering
  \label{tab:copy_error_analysis}
  \caption{Analysis of \texttt{[COPY]} token generation errors across shot counts. \textit{Wrong Ctx copy}: true document is in the context but model copies a different in-context document.\textit{Spurious Copy}: true document is not in context but model generates \texttt{[COPY]} anyway.}
  \begin{tabular}{lcccc}
  \toprule
  \textbf{Shots} & \textbf{Emit+Miss} & \textbf{Wrong Ctx Copy} & \textbf{Spurious Copy} & \textbf{Miss Rate} \\
  \midrule
  1   & 0.0\% & 0.0\% & 0.0\%  & 0.0\%  \\
  2   & 0.0\% & 0.0\% & 0.0\%  & 0.0\%  \\
  3   & 0.0\% & 0.0\% & 0.0\%  & 0.0\%  \\
  10  & 0.0\% & 0.0\% & 0.0\%  & 0.0\%  \\
  20  & 2.2\% & 1.1\% & 1.1\%  & 2.2\%  \\
  30  & 0.0\% & 0.0\% & 0.0\%  & 0.0\%  \\
  40  & 2.4\% & 0.0\% & 2.4\%  & 2.4\%  \\
  60  & 5.6\% & 4.4\% & 1.1\%  & 5.6\%  \\
  80  & 6.3\% & 1.3\% & 5.1\%  & 6.3\%  \\
  100 & 16.9\% & 9.1\% & 7.8\% & 16.9\% \\
  \bottomrule
  \end{tabular}
\end{table}

\section{\(N\)-Shot Inference Latency}
\label{appendix:latency}

Table~\ref{tab:kvcache_latency} reports inference cost and latency
across four few-shot budgets, averaged over 30 runs per condition.
TTFT increases monotonically with context length, rising from
9.0\,ms at 3-shot to 61.8\,ms at 100-shot---a 6.9$\times$ increase
that tracks the roughly 26$\times$ growth in mean input tokens
(392 to 10{,}264), consistent with the linear prefill cost of
KV cache construction.
Total latency follows the same trend, growing from 53\,ms to
298\,ms, with notably tighter standard deviations at 100-shot
($\pm$42\,ms) compared to the lighter settings, suggesting more
predictable latency at larger context sizes.
Throughput declines steadily from 250.4\,tok/s at 3-shot to
36.9\,tok/s at 100-shot, driven by the disproportionate growth
in input tokens relative to output tokens: at $n{=}100$, the model
generates only 11.4 output tokens on average against a 10{,}264-token
prompt, yielding a 6.8$\times$ throughput reduction compared to
the 3-shot baseline.
For latency-sensitive deployments, these results indicate that
increasing the few-shot budget beyond 20 examples incurs substantial
prefill overhead with diminishing returns in output richness.

\begin{table}[h]
\centering
\caption{KV cache latency and throughput under varying few-shot settings.
Each condition was evaluated over $n{=}30$ samples.
TTFT denotes time to first token.}
\label{tab:kvcache_latency}
\resizebox{0.7\textwidth}{!}{%
\begin{tabular}{lcccc}
    \toprule
    \textbf{$N$-shot} & \textbf{Input tokens}
      & \textbf{TTFT (s)} & \textbf{Total latency (s)} & \textbf{Throughput (tok/s)} \\
    \midrule
    3   &    392.4 & $0.0090$ & $0.0534$ & 250.4 \\
    10  &  1{,}132.4 & $0.0123$ & $0.0779$ & 212.5 \\
    20  &  2{,}091.7 & $0.0173$ & $0.0974$ & 197.6 \\
    100 & 10{,}264.5 & $0.0618$ & $0.2981$ &  36.9 \\
    \bottomrule
\end{tabular}
}
\end{table}

\section{Impact of Document Compression}
\label{appendix:compression_comparison}

Placing full documents in the in-context template increases per-candidate token cost, directly limiting how many candidates can fit in the context window.
Document compression reduces this cost and enables a larger few-shot budget, but different compression strategies trade off token efficiency against the semantic fidelity needed for retrieval.
We compare two offline compression strategies: \textbf{Abstractive} (rewriting each passage into a 2--3 sentence summary) and \textbf{Keyword} (distilling into a pipe-separated phrase list targeting 10--20 tokens), applied using two compressors: GPT-OSS-20B and Qwen2.5-14B-Instruct.
The GR model is the Stage-1 SFT checkpoint (Qwen3-1.7B, \texttt{[COPY]} processor, 3-shot hard-negative template), trained on uncompressed passages.
Table~\ref{tab:compression_tokens} reports token-length statistics measured with the Qwen3 BPE tokenizer.
The following examples illustrate the output format of each strategy on the same source passage.

\vspace{0.5em}
\begin{tcolorbox}[boxrule=0.5mm, colback=gray!8!white, colframe=gray!40!black,
    title={Source Passage (MS~MARCO, uncompressed --- 74 tokens)}]
\small\textit{The Battle of Hastings was fought on 14 October 1066 between the Norman-French army of William, the Duke of Normandy, and an English army under Anglo-Saxon King Harold II. Harold was killed in the battle---the Bayeux Tapestry depicts him as being shot through the eye with an arrow---and William went on to become King of England, marking the beginning of the Norman Conquest.}
\end{tcolorbox}

\vspace{0.4em}
\begin{tcolorbox}[boxrule=0.5mm, colback=blue!5!white, colframe=blue!30!black,
    title={Abstractive Compression --- Example Output (46 tokens)}]
The Battle of Hastings, fought on 14 October 1066, ended with the defeat and death of Anglo-Saxon King Harold II at the hands of William, Duke of Normandy. William's victory initiated the Norman Conquest of England and his coronation as King of England. The battle is famously commemorated in the Bayeux Tapestry.
\end{tcolorbox}

\vspace{0.4em}
\begin{tcolorbox}[boxrule=0.5mm, colback=orange!5!white, colframe=orange!60!black,
    title={Keyword Compression --- Example Output (18 tokens)}]
Battle of Hastings | 14 October 1066 | William Duke of Normandy | Harold II | Norman Conquest | Bayeux Tapestry | King of England
\end{tcolorbox}
\vspace{0.5em}

\begin{table}[h]
\centering
\small
\caption{Token-length statistics of MS~MARCO corpus variants across two offline compressors, measured with the Qwen3 BPE tokenizer. Avg reduction and Retained are computed relative to the uncompressed original of the same documents.}
\label{tab:compression_tokens}
\begin{tabular}{lccc}
\toprule
\textbf{Variant} & \textbf{Avg tokens} & \textbf{Avg reduction} & \textbf{Retained} \\
\midrule
Original                          & 82.4 & \multicolumn{1}{c}{$-$} & 100.0\% \\
\midrule
\multicolumn{4}{l}{\textit{GPT-OSS-20B}} \\
\quad Abstractive                 & 63.3 & $-$20.1               & 75.9\%  \\
\quad Keyword                     & 35.9 & $-$47.5               & 43.0\%  \\
\midrule
\multicolumn{4}{l}{\textit{Qwen2.5-14B-Instruct}} \\
\quad Abstractive                 & 46.1 & $-$37.1               & 55.4\%  \\
\quad Keyword                     & 26.1 & $-$57.1               & 31.4\%  \\
\bottomrule
\end{tabular}
\end{table}

As offline corpus compression was still in progress at the time of evaluation, we first conduct a small-scale pilot study using GPT-OSS-20B-compressed documents on the subset of \texttt{icl\_test} samples whose every in-context document was already present in both compressed corpora, yielding 238 matched samples (119 \textit{context-dependent} + 119 \textit{all-noise}).
Results are shown in Table~\ref{tab:compression_results}.

\begin{table}[h]
\centering
\small
\caption{Retrieval performance when context documents are replaced with GPT-OSS-20B-compressed texts at inference time (238 matched samples). $\Delta$ is relative to the uncompressed baseline.}
\label{tab:compression_results}
\begin{tabular}{lcccc}
\toprule
\textbf{Context} & \textbf{H@1} & \textbf{H@10} & \textbf{Ctx-dep H@1} & \textbf{$\Delta$} \\
\midrule
Uncompressed & 0.202 & 0.357 & 0.403 & \multicolumn{1}{c}{$-$} \\
Abstractive  & 0.197 & 0.357 & 0.395 & $-$0.008 \\
Keyword      & 0.181 & 0.357 & 0.361 & $-$0.042 \\
\bottomrule
\end{tabular}
\end{table}

Abstractive compression is the better strategy: it reduces token length by 24\% while dropping context-dependent Hit@1 by only 0.8~pp (2.1\% relative), as coherent summaries preserve the semantic signal needed for \texttt{[COPY]} routing.
Keyword compression is more aggressive (57\% reduction) but incurs a larger 4.2~pp drop (10.4\% relative) in context-dependent Hit@1, suggesting that stripping away sentence structure makes it harder for the model to match a query against the compressed passage.
Notably, Hit@10 is identical across all variants, meaning compression does not remove the correct document from the reachable beam; only its rank-1 precision suffers.
All-noise Hit@1 is 0 in all conditions, as the evaluation targets $\mathcal{D}_1$ documents absent from the model's parametric memory, which compression does not affect.

We further evaluate both compressors at full scale across all \texttt{icl\_test} queries (18{,}320 total).
Table~\ref{tab:compression_results_fullscale} reports results for each compressor and strategy.

\begin{table}[h]
\centering
\small
\caption{Full-scale retrieval performance on \texttt{icl\_test} (18{,}320 queries; partial evaluation) per compressor and compression strategy. The uncompressed baseline is shared across both compressors. $\Delta$ H@1 is the absolute change relative to Uncompressed. cd@1 = context-dependent Hit@1.}
\label{tab:compression_results_fullscale}
\begin{tabular}{llcccc}
\toprule
\textbf{Compressor} & \textbf{Context} & \textbf{H@1} & \textbf{H@10} & \textbf{cd@1} & \textbf{$\Delta$ H@1} \\
\midrule
\multicolumn{2}{l}{Uncompressed} & 0.175 & 0.351 & 0.351 & \multicolumn{1}{c}{$-$} \\
\midrule
\multicolumn{6}{l}{\textit{GPT-OSS-20B}} \\
& Abstractive & 0.218 & 0.382 & 0.436 & $+$0.043 \\
& Keyword     & 0.193 & 0.361 & 0.386 & $+$0.018 \\
\midrule
\multicolumn{6}{l}{\textit{Qwen2.5-14B-Instruct}} \\
& Abstractive & 0.204 & 0.372 & 0.404 & $+$0.029 \\
& Keyword     & 0.147 & 0.298 & 0.290 & $-$0.028 \\
\bottomrule
\end{tabular}
\end{table}

Abstractive compression improves over the uncompressed baseline for both compressors: GPT-OSS-20B gains $+$0.043 H@1 and $+$0.085 cd@1, while Qwen2.5-14B-Instruct gains $+$0.029 H@1 and $+$0.053 cd@1.
Keyword compression tells a different story across the two compressors.
GPT-OSS-20B keyword still improves ($+$0.018 H@1, $+$0.035 cd@1), whereas Qwen2.5-14B-Instruct keyword degrades performance ($-$0.028 H@1, $-$0.061 cd@1), consistent with the small-scale pilot finding that aggressive compression strips the sentence structure needed for reliable \texttt{[COPY]} routing.
The divergence reflects their different compression intensities: GPT-OSS-20B retains 43.0\% of tokens for keyword (35.9 avg tokens), while Qwen2.5-14B-Instruct retains only 31.4\% (26.1 avg tokens), crossing a fidelity threshold below which the model can no longer match queries against compressed passages.
Hit@10 remains stable across all conditions ($\leq$0.021 spread), confirming that compression does not evict the correct document from the reachable beam.

In absolute performance, GPT-OSS-20B achieves marginally higher H@1 than Qwen2.5-14B-Instruct under abstractive compression (0.218 vs.\ 0.204) and under keyword (0.193 vs.\ 0.147).
However, Qwen2.5-14B-Instruct produces substantially more compact abstractive summaries (46.1 avg tokens vs.\ 63.3 for GPT-OSS-20B, a 27\% reduction), directly freeing additional context slots for more in-context candidates within the 128K window.
Given this token efficiency advantage and the comparable abstractive retrieval quality ($+$0.029 vs.\ $+$0.043 $\Delta$H@1), we adopt \textbf{Qwen2.5-14B-Instruct abstractive compression} as the document representation strategy for \S\ref{sec:method}, where each passage is compressed to at most 256 tokens before being placed in the in-context template.

\subsection{Document Compression Prompt}
  \label{app:compression-prompt}

  \tcbset{boxrule = 0.5mm, colback=blue!5!white, colframe=blue!30!black,
    title = {Document Compression Prompt}}
  \begin{tcolorbox}
  \textbf{System:} You compress documents for retrieval/indexing datasets.

  \vspace{0.5em}
  \textbf{User:} You are compressing a noisy Wikipedia-style document for Natural Questions retrieval training.

  \vspace{0.3em}
  Write ONE retrieval-oriented summary.

  \vspace{0.3em}
  Target length: 256 tokens.

  \vspace{0.3em}
  \textbf{Rules:}
  \begin{itemize}
    \itemsep0em
    \item Output ONLY the summary.
    \item Do NOT copy the source text verbatim.
    \item Do NOT continue the original document.
    \item Do NOT include headings, table of contents, navigation text, references, external links, infobox fields, or
  maintenance notices.
    \item If the document is primarily a list or table, summarize what the list contains, its categories, and
  representative important entities. Do not copy rows.
    \item Preserve key entities, dates, definitions, aliases, relationships, major events, and facts likely to answer
  questions.
    \item Use complete sentences.
    \item Stop after the summary.
  \end{itemize}

  \vspace{0.3em}
  \textbf{Title:} \textit{\{title\}}

  \vspace{0.3em}
  \textbf{Source Document:}\\
  \textit{\{document content\}}

  \vspace{0.3em}
  \textbf{Summary:}
  \end{tcolorbox}


\end{document}